\documentclass[sigchi]{acmart}

\setcopyright{acmlicensed}
\copyrightyear{2018}
\acmYear{2018}
\acmDOI{XXXXXXX.XXXXXXX}

\acmConference[Conference acronym 'XX]{Make sure to enter the correct
  conference title from your rights confirmation email}{June 03--05,
  2018}{Woodstock, NY}

\acmISBN{978-1-4503-XXXX-X/2018/06}

\usepackage{amsmath}
\usepackage{graphicx}
\usepackage{booktabs}
\usepackage[most]{tcolorbox}
\usepackage{fvextra}
\usepackage{listings}
\usepackage{xcolor}

\definecolor{guideblue}{HTML}{0D90B2}
\definecolor{guideghreen}{HTML}{19A249}
\definecolor{guidepurple}{HTML}{9234EA}
\definecolor{guidelightblue}{HTML}{D0FAFE}
\definecolor{guidepink}{HTML}{DB2876}
\definecolor{guidelightgreen}{HTML}{DBFCE7}
\definecolor{guideorange}{HTML}{EA570B}
\definecolor{guidelightpurple}{HTML}{F3E8FF}
\definecolor{guidelightpink}{HTML}{FDE6F4}
\definecolor{guidelightorange}{HTML}{FFECD5}

\newtcblisting{promptbox}{
    breakable,
    enhanced,
    colback=gray!5,
    colframe=gray!40,
    boxrule=0.4pt,
    arc=1.5pt,
    left=4pt,
    right=4pt,
    top=3pt,
    bottom=3pt,
    listing only,
    listing options={
        breaklines=true,
        breakatwhitespace=false,
        basicstyle={\scriptsize\ttfamily\linespread{0.85}\selectfont},
        xleftmargin=0pt,
        xrightmargin=0pt
    }
}

\begin{document}

\title{GUIDE: Designer-in-the-loop Authoring of Conformant Generative User Interfaces}

\author{Hyewon Lee}
\email{hyewon@purdue.edu}
\affiliation{%
  \institution{Department of Computer Science, Purdue University}
  \city{West Lafayette, Indiana}
  \country{USA}
}

\author{Ziying Wang}
\email{ziying.wang@dukekunshan.edu.cn}
\affiliation{%
  \institution{Duke Kunshan University}
  \city{Kunshan, Jiangsu}
  \country{China}
}

\author{Aiden Moy}
\email{moy32@purdue.edu}
\affiliation{%
  \institution{Department of Computer Science, Purdue University}
  \city{West Lafayette, Indiana}
  \country{USA}
}

\author{Saran Nagubandi}
\email{snaguban@purdue.edu}
\affiliation{%
  \institution{Department of Computer Science, Purdue University}
  \city{West Lafayette, Indiana}
  \country{USA}
}

\author{Jason Wu}
\email{jasonwu@purdue.edu}
\affiliation{%
  \institution{Department of Computer Science, Purdue University}
  \city{West Lafayette, Indiana}
  \country{USA}
}


\begin{abstract}
Generative User Interfaces (GenUIs) enable applications to generate interfaces on demand from user needs and context. Like conventional UIs, they must still reflect designers' intent and conform to requirements such as brand identity. Unlike conventional UIs, designers cannot directly specify or see every interface a GenUI may produce, making design intent harder to enforce.
We introduce \textbf{GUIDE (GenUI Development Environment)}, a system that lets designers continuously inspect and refine GenUI behavior as they create and edit interfaces. GUIDE uses designers' modifications and interactions to adapt GenUIs through prompt optimization and a novel adaptive conformance scoring model.
We validate GUIDE's scoring model and system. The scoring model matched or outperformed proprietary LLM baselines on held-out comparisons of real and synthetic application screens. In a study with 12 UI/UX practitioners, participants found GUIDE effective and usable and significantly preferred aligned outputs over a strong baseline using exemplars and a model-generated \texttt{design.md}.
\end{abstract}

\begin{teaserfigure}
  \centering
  \includegraphics[width=\textwidth]{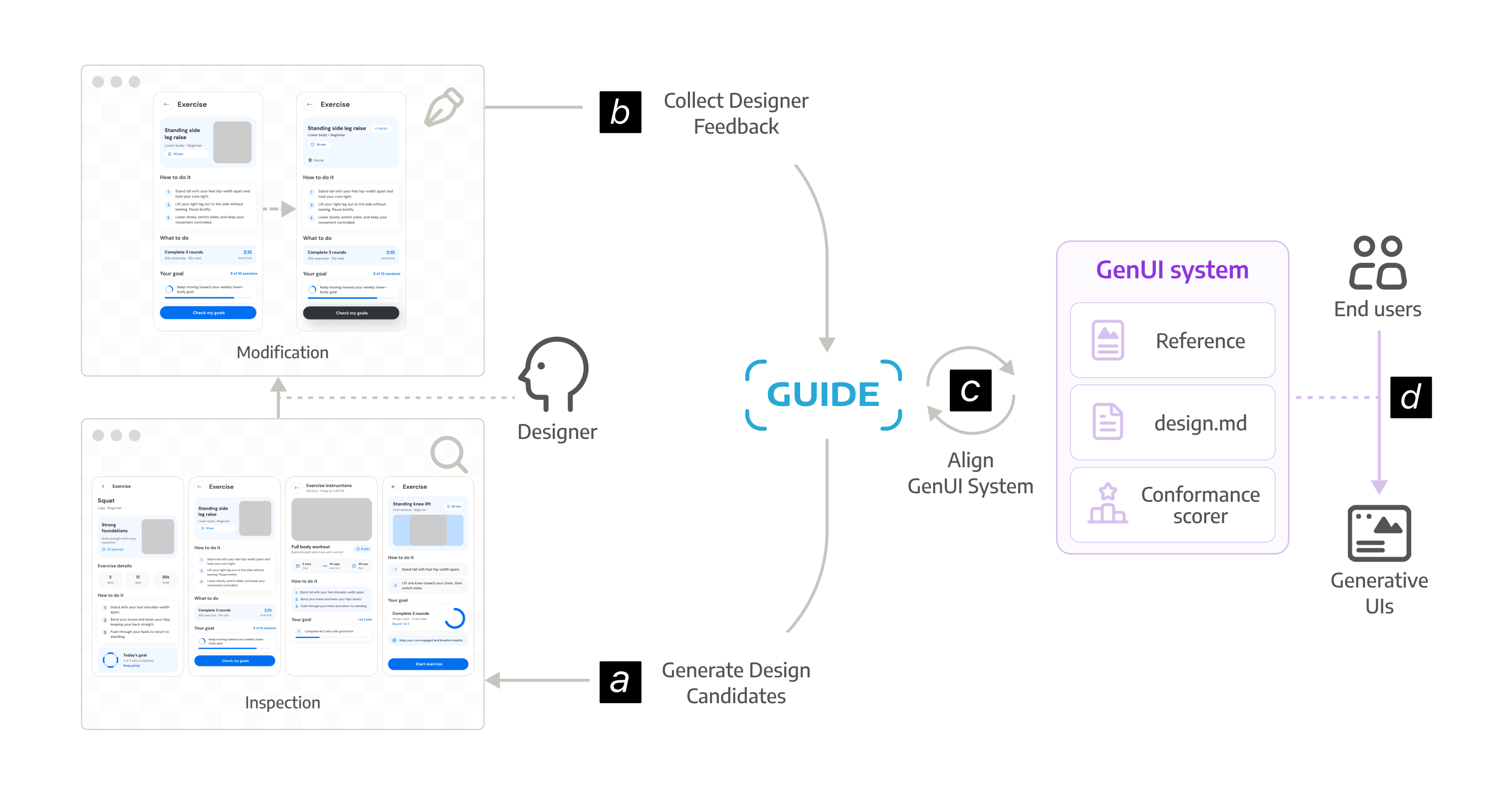}
  \caption{Architectural overview of GUIDE, illustrating a designer-in-the-loop workflow for aligning a GenUI system. GUIDE leverages the GenUI system, which consists of a reference set, \texttt{design.md}, and a conformance scorer, to (a) generate design candidates. The designer inspects and modifies one screen to better match their intent. (b) The collected designer feedback is used to (c) update each component of the GenUI system. This iterative cycle allows the designer to steer the GenUI system to produce (d) conformant generative UIs for end users.}
  \Description{A left-to-right diagram of the GUIDE workflow. Six existing Figma artboards enter GUIDE, which produces three inputs: a designer prompt, a \texttt{design.md} file containing design-system properties, and visual references. A generator uses these inputs to produce 32 candidate interfaces. A scoring model, trained on a separate and visually diverse UI dataset, ranks the candidates and selects the best variant. The selected interface is presented in the Figma plugin for designer editing. The resulting edits become fine-tuning data that feeds back into both the generator and the scoring model.}
  \label{fig:teaser}
\end{teaserfigure}
\maketitle

\section{Introduction}
Unlike conventional user interfaces (UIs) which are manually designed and implemented, Generative User Interfaces (GenUIs) generate interfaces on demand from user needs and context. While this capability has great potential for end-users, it also raises a new challenge for designers: \textit{how do you design an interface that does not yet exist?} Traditional design tools let designers explore and refine concrete interfaces before deployment, but for GenUIs, the interface being refined is only one possible output of a system that will generate others at runtime. 

This suggests that in order to effectively design GenUIs, designers need generalizable ways to shape the system's behavior beyond individual interfaces.
Prior work has explored several approaches, such as toolkits that encode design constraints~\cite{churchill2019scaling,min2025meridian}, prompting~\cite{cao2025generative,park2026bridging}, and learned evaluators that steer generation or selection~\cite{wu2024uiclip,wu2026improving}.
While these approaches can be used to alter the system's output, they do not by themselves tell designers whether the system's broader behavior matches their intent. Designers can author examples or write guidelines, but these only represent pieces of their full notion of conformance, which is tacit and hidden from the system~\cite{son2024demystifying,park2026bridging}.
At the same time, the system's actual interpretation (e.g., full output distribution) built from these partial resources is also hidden from the designer except through a sparse set of generated samples, which makes it difficult for designers to know what part of that interpretation needs to be adjusted.

In this paper, we introduce \textbf{GUIDE (GenUI Development Environment)}, a system implemented as a Figma plugin for designer-in-the-loop alignment of GenUIs. As the name suggests, GUIDE allows designers to guide a GenUI towards their design intent by continuously inspecting and refining output using canvas edits, which are used by the plugin to optimize a GenUI prompt and tune a project-specific scoring model. While prior work has also shown how natural designer interactions can be converted into supervision for UI generation models~\cite{wu2026improving}, GUIDE brings this learning directly into the design loop. Designers can then test these updates through subsequent generations, using each iteration to target remaining mismatches between the system's interpretation and their intent.
To build our system, we first constructed a dataset of 2,071 real-world UI designs from 52 design systems~\cite{chai2025amex}, augmented with intra-app references and synthetic augmentations to capture variation within and across design systems. We used this dataset to develop a generalizable design conformance model that evaluates candidate designs against a learned style embedding that can be computed from a reference set of screenshots and other types of designer feedback. As designers work with GUIDE, their artboards and refinements are then used to optimize an input prompt and adapt the conformance model to the current project.

We validate our approach in several ways. To evaluate our design conformance model, we conducted a technical evaluation on held-out applications from our dataset, testing whether its learned representation separates design systems and its prediction accuracy for known conformance relationships. We found that our local model significantly outperformed large state-of-the-art models (GPT-5.6 Luna) by up to 11\% in design conformance prediction. To evaluate the usability and effectiveness of the GUIDE system, we conducted a user study with 12 participants with design experience who used GUIDE to align GenUIs for artboards found on the Figma Community repository. All participants rated GUIDE as highly usable and in a blind rating task, they significantly preferred designs generated by their aligned GenUIs over those generated by a strong baseline using in-context examples and a model-generated \texttt{design.md}.

To summarize, our paper makes the following contributions:
\begin{enumerate}
    \item We contribute a computational formulation of design conformance, which we operationalize into a dataset and model.
    \item We contribute the GUIDE system, which integrates our design conformance model and a prompt optimization technique to enable designers to interactively align a GenUI's output.
\end{enumerate}

To facilitate research in this area, we plan to open source our models and datasets upon acceptance.

\section{Related Work}
To contextualize our work on conformant GenUI systems, we review three related areas: (i) UI generation, (ii) methods for expressing design intent, and (iii) generative alignment. 

\subsection{UI Generation}

HCI and ML have long studied how to automatically generate user interfaces. Early model-based UI approaches (MBUI) typically took a task model and a domain model as input and derived a corresponding interface representation from them~\cite{puerta1998towards,gajos2004supple}. For example, Puerta et al. described model-driven frameworks for supporting cross-platform, reusable interface development~\cite{puerta1998towards}, and SUPPLE formulated interface generation as an optimization problem, searching among candidate interfaces against user-specific ability models~\cite{gajos2004supple}.

The extensive modeling requirements of MBUIs made them challenging to apply in domains that are difficult to formalize. Consequently, later work explored generating interface layouts using lighter-weight inputs. For example, systems utilizing constraint and optimization algorithms like DesignScape, GRIDS, Sketchplore, and Scout recommend layouts by translating structural rules such as alignment and visual hierarchy into objective functions~\cite{o2015designscape,dayama2020grids,todi2016sketchplore,swearngin2020scout}. A separate line of work learns layout distributions from data: LayoutGAN models geometric relationships among elements, LayoutTransformer learns contextual relations through self-attention, and PLay uses diffusion to generate layouts conditioned on designer-provided guidelines~\cite{li2020layoutgan,gupta2021layouttransformer,cheng2023play}. Related data-driven systems have also learned page and poster structure directly from examples~\cite{qiang2016learning}. Together, this body of work focuses primarily on the visual structure and spatial organization of interface elements.

Large language models have further expanded the scope of interface generation, enabling systems to generate executable interface code directly from natural language descriptions. Earlier learning-based systems had already explored translating interface representations into implementation code, including screenshot-to-code approaches such as pix2code~\cite{beltramelli2018pix2code}. General-purpose LLMs substantially broaden this capability~\cite{achiam2023gpt}, while recent systems combine language models with interface representations, retrieval, and iterative prototyping to support increasingly complete UI generation workflows~\cite{leviathan2026generative,petridis2023promptinfuser,yuan2026towards}. Researchers have also developed models specifically for UI code generation; UICoder uses compiler and vision-language-model feedback to train models to produce compilable interface code without manually annotated feedback~\cite{wu2024uicoder}.

Despite improving generation capabilities, a gap remains between GenUI tools and designers' actual needs~\cite{chen2025genui,park2026bridging}. Professional designers still struggle using prompt-based GenUI tools to articulate intent, interpret results, and steer revisions~\cite{chen2025genui,park2026bridging,zamfirescu2023johnny}. Together, these findings suggest that supporting designers in continuously expressing and adjusting design requirements over the course of generation remains an open problem. Our work follows this direction, letting designers guide multi-round generation through selection, editing, and feedback.

\subsection{Expressing Design Intent}

Interface designs often originate in free-form tools, requiring translation into executable behavior during implementation. Early toolkits captured this intent using programming abstractions and constraint systems. For instance, Amulet supports interface constraints and animations, while Cassowary expresses spatial relationships through linear constraints~\cite{myers1997amulet, badros2001cassowary}. While these approaches allow systems to execute design requirements directly, they force designers to explicitly encode them as programs, rules, or constraints.

HCI has also long explored expressing interface behavior through demonstration, forming a broader line of work on demonstrational interfaces and programming by demonstration~\cite{myers1990creating, lau1998programming, mcdaniel1998building, cypher1993watch, lieberman2001your, gulwani2011automating, li2017sugilite, chasins2018rousillon}. Pavlov lets designers construct interface behavior by demonstrating the relationship between stimuli and responses~\cite{wolber1997pavlov}, while FrameKit supports authoring adaptive interfaces through keyframe demonstration~\cite{wu2024framekit}. Related work on example-based specification similarly shows that concepts that are difficult to state abstractly can become more concrete when grounded in examples~\cite{li2019pumice}. Machine teaching and interactive machine learning extend this idea by treating examples, corrections, and interactive feedback as ways for people to communicate task knowledge to a learning system~\cite{zhu2015machine, simard2017machine, ramos2020interactive, fails2003interactive, amershi2014power}, including subjective domains in which people judge outcomes holistically rather than through fully specified rules~\cite{fiebrink2011human, fogarty2008cueflik}.

Design intent can also be distributed across a team, an organization, and the artifacts produced over the course of a design process. Design systems record and maintain shared design knowledge through components, interaction patterns, visual specifications, and usage rules, and support its ongoing transmission across products and team members~\cite{lamine2022understanding, feng2023understanding}. Similarly, design manuals and style guides~\cite{lamine2022understanding} and reference screens~\cite{klemmer2002web} can all carry explicitly recorded design requirements. At the same time, part of this design knowledge remains tacit: design judgment has long been characterized as depending on knowledge that practitioners can apply without fully articulating it~\cite{polanyi2009tacit}, and Son et al. show that such tacit knowledge in design practice often depends on experience, judgment, and specific context~\cite{son2024demystifying}. Related work further shows that activity traces, such as selections, edits, iterations, annotations, and references, can preserve design process information that is never explicitly recorded~\cite{hammad2026tracing}.

Our work builds on this body of research, converting the explicit inputs, selections, and edits a designer produces during generation into computational representations that GenUI can continuously use.

\subsection{Generative Alignment}

The content and behavior of a generative interface are produced dynamically by a generative model, so a design target must be further translated into a signal that can constrain the model's output. Model alignment research has broadly explored how to shape a generative model's behavior through input optimization, human feedback, and learned reward signals~\cite{christiano2017deep, ouyang2022training, nakano2021webgpt, stiennon2020learning, khattab2023dspy, yuksekgonul2024textgrad}. One path directly relevant to generative design is optimizing the prompt or other input a model receives. Generative models are highly sensitive to how an input is phrased, and even when a user knows what they want, they often struggle to translate that goal into a prompt that reliably produces it~\cite{zamfirescu2023johnny, li2019pumice}. HCI systems help users refine prompts through interactive feedback: APPO optimizes prompts based on user choices among candidate results, while PromptInfuser couples authoring with iterative UI refinement~\cite{li2026preference, petridis2023promptinfuser}. ML work instead formalizes prompt search more explicitly as an optimization problem: DSPy optimizes the prompts and demonstrations within an LLM pipeline against a task metric, while TextGrad and GEPA revise textual input through natural language feedback, execution traces, and reflection~\cite{khattab2023dspy, yuksekgonul2024textgrad, agrawal2026gepa}.

A second line of approaches shape model behavior by evaluating its generated outputs, either during training or inference. Preference learning and RLHF derive reward signals from human comparisons and use them to rank, select, or optimize model outputs~\cite{christiano2017deep, stiennon2020learning, nakano2021webgpt, ouyang2022training}. HCI research has further explored how such evaluation objectives can be formed from user behavior and the specific context of use. Just-in-time objectives infer a goal from a user's current behavior and apply it to subsequent generation and evaluation~\cite{lam2026just}, while PosterMate provides structured feedback on a design through the evaluative perspectives of multiple personas~\cite{shin2025postermate}. Related work in visual generation similarly learns scoring functions from human judgments to rank candidate images or generated outputs~\cite{xu2023imagereward, kirstain2023pick, wu2023human}. This body of work converts user preferences, in-the-moment behavior, and task-specific criteria into computational signals that can be used to evaluate and guide generation.

For interface and visual design specifically, work has further learned evaluation models tailored to the domain. Earlier work predicts visual or aesthetic quality directly from images or relative comparisons~\cite{murray2012ava, reinecke2013predicting, kong2016photo}, while UIClip learns design-relevant correspondence between interface visuals and text for evaluating and ranking generated interfaces~\cite{wu2024uiclip}. Other work evaluates conformance to a particular interface or design language, including detecting differences between an implementation and its design mockup, modeling adherence to a visual design language, and measuring consistency across screens within a product~\cite{moran2018automated, doosti2018computational, park2023computational}. DesignPref further models differences in individual designers' preferences, allowing an evaluation signal to be personalized to a specific user~\cite{peng2026efficient}. Closest to our work is Wu et al.~\cite{wu2026improving}, who convert designer feedback (e.g., comments, sketches) into training data to improve generation models. While they focus on offline model training, our work enables designers to continuously test and adjust alignment during live generation, simultaneously updating prompt optimization and scoring mechanisms.
\section{Conceptualizing Design Conformance}

\begin{figure}[t]
    \centering
    \includegraphics[width=\columnwidth]{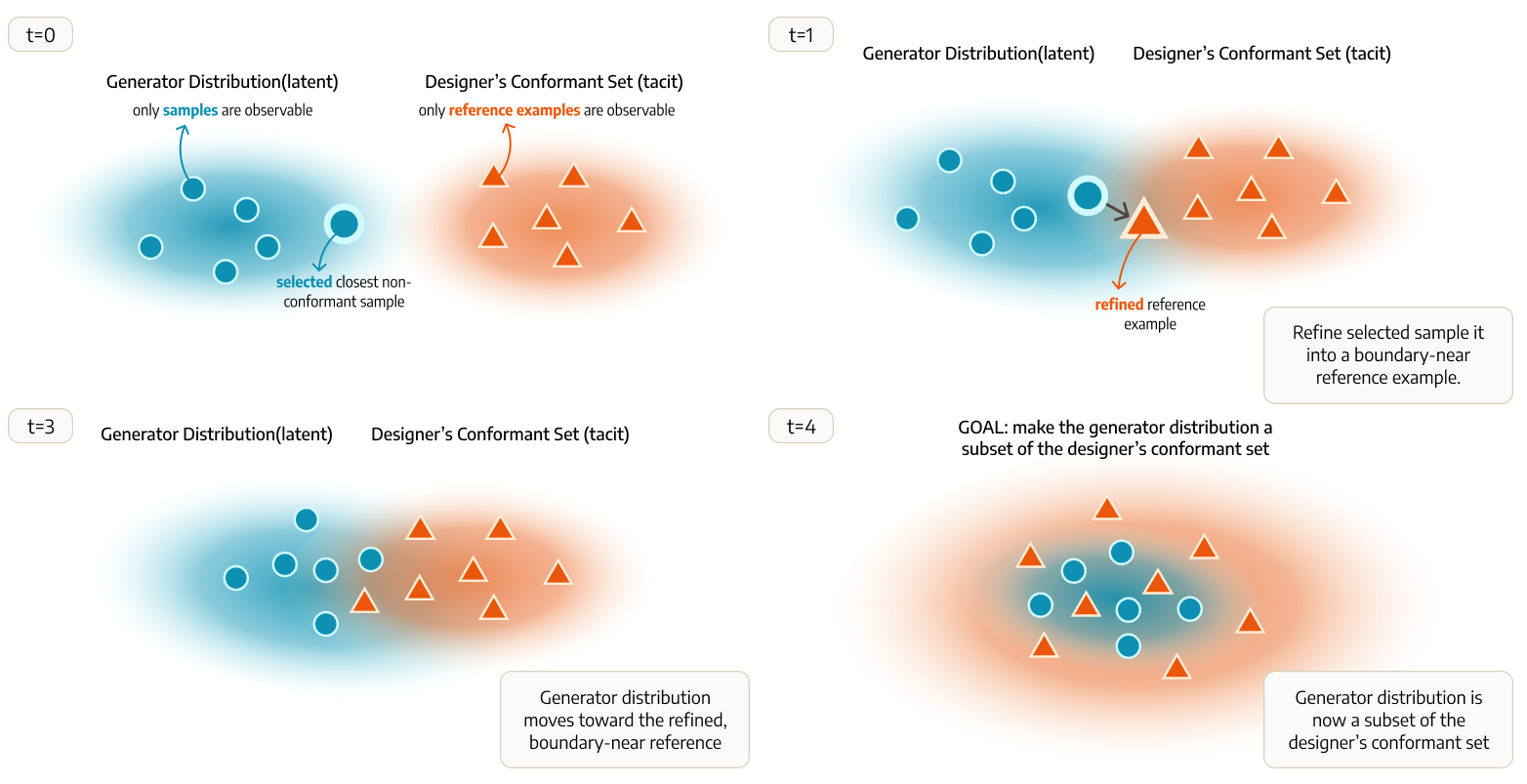}
    \caption{A conceptualization of the iterative process of aligning a GenUI system. The generator's latent output distribution (blue) is shifted toward the designer's tacit conformant set (orange) through designer guidance, eventually making the generator's output distribution a subset of the conformant set.}
    \Description{This four-part diagram shows the alignment of a GenUI system by shifting the generator's output distribution (a blue cloud with circular samples) toward the designer's conformant set (an orange cloud with triangular examples). At t=0, the sets are separate, and the blue sample closest to the orange set is selected. At t=1, this sample is refined into a boundary-near reference, moving into the orange set as a blue-bordered triangle. By t=3, the entire blue distribution shifts to heavily overlap the orange set, drawn toward the refined reference. Finally, at t=4, the goal is reached: the blue generator distribution is fully contained as a subset within the orange conformant set.}
    \label{fig:conformance-concept}
\end{figure}

In this paper, we use \textit{design conformance} to describe how well a design matches a designer's intent, which may include conforming to a brand's design language. Unlike general interface quality, conformance is specific to that intent, so even a polished and usable interface can be non-conformant. At the same time, many different interfaces can conform to the same intent, and a designer may ultimately select one based on other factors, such as personal preference. Finally, designers may disagree about whether a design conforms, as well as how closely it matches the intended design goal~\cite{churchill2019scaling}.

In practice, designers communicate design intent through design systems, documentation, toolkit implementations, and reference designs. Some of these resources specify concrete properties that can be enforced directly, as prior computational design systems have done with layout relationships~\cite{dayama2020grids}. Yet restricting a design to use only approved components or a toolkit that enforces a set of rules does not guarantee conformance.
For example, limiting a design to approved components can rule out acceptable customized widgets while still allowing those components to be combined in ways that is inconsistent with other screens in the application.
Ultimately, designers rely on tacit knowledge of the product's design intent to determine the boundaries of acceptable variation~\cite{son2024demystifying}.

\subsection{Conformance in Generative User Interfaces}

Enforcing design conformance in GenUIs raises unique challenges. Unlike conventional design workflows where designers review each artifact directly, GenUIs require designers to assess conformance from sampled outputs that may differ from the interfaces later generated for end users.

In this paper, we model a GenUI as a stochastic generator whose output distribution is shaped by its input, learned parameters, and inference algorithm (e.g., rejection sampling). Given a dynamically constructed prompt \(x\), a GenUI \(G\) generates an interface \(y \sim G(x)\). Let \(c(y)\) denote the designer's latent conformance judgment for \(y\). Aligning the GenUI means shifting \(G(x)\) toward outputs with higher \(c(y)\).
Designers may try to shape \(G\) through prompt templates, guidelines, or examples that communicate their design intent. Yet many aspects of design are difficult to articulate~\cite{park2026bridging}, and even a complete specification may not be interpreted by the generator as intended. The designer can observe \(G(x)\) only through generated instances, while the GenUI can observe \(c(y)\) only through what the supplied artifacts make explicit.

Having a designer-in-the-loop is essential because it ties output samples from \(G(x)\) directly to judgments of \(c(y)\) by gradually revealing the generator's behavior to the designer and the designer's intent to the generator. Because feedback is elicited on outputs from the current generator, designer judgment is directed toward errors that actually arise from a model's mis-aligned interpretation. When a designer edits a nearly acceptable output into one they would accept, the before-and-after pair shows which specific changes separate the two judgments. Finally, as \(G\) ``narrows in'' on the designers' tacit intent, later generations can surface remaining ambiguities for the designer to resolve. Figure \ref{fig:conformance-concept} provides a visualization for our conceptualization of this process.
\begin{tcolorbox}[
    colback=guidelightblue!35!white,
    colframe=guideblue!20!white,
    boxrule=0.4pt,
    left=6pt,
    right=6pt,
    top=4pt,
    bottom=4pt,
    arc=1pt
]
A GenUI is conformant when its output distribution aligns with a designer's judgments of which interfaces fit their intent. Designers can observe only samples of the generator's behavior, while the GenUI has access only to the parts of designer intent made explicit through artifacts and feedback. Keeping the designer in the loop allows both the designer and GenUI to gradually reveal the feedback needed for alignment.
\end{tcolorbox}

\section{Learning Design Conformance from Data}

\begin{figure*}[!htb]
    \centering
    \includegraphics[width=0.8\textwidth]{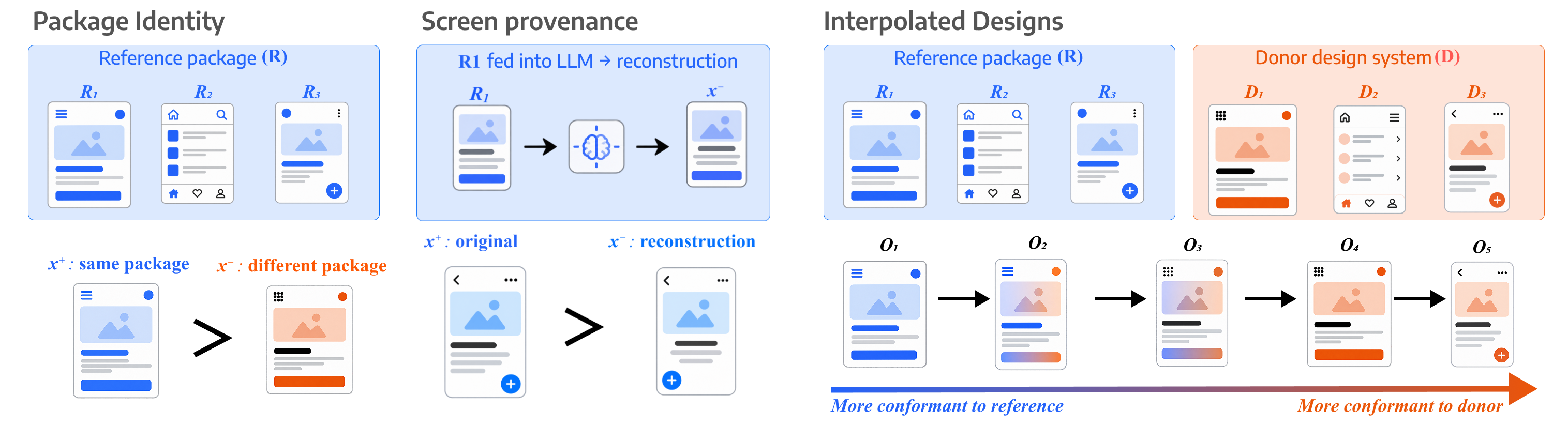}
    \caption{Overview of our dataset generation pipeline that generates three types of conformance relations using the package identities of real screenshots (Left), whether a screen was a real screenshot or generated (Center), and interpolations of synthetically generated screens (Right).}
    \Description{This diagram details the dataset generation strategies across three panels. Our dataset generation pipeline generates three types of conformance relations using the package identities of real screenshots (Left), whether a screen was a real screenshot or generated (Center), and interpolations of synthetically generated screens (Right).}
    \label{fig:data-generation}
\end{figure*}
To operationalize this view of design conformance, we first construct a dataset that captures variation within design systems and examples of generated designs with varying levels of conformance to them. We then use these relationships to explore methods for learning design conformance computationally.

\subsection{Dataset Construction}

We used the AMEX dataset as a large-scale source of real-world UI designs, with multiple screens from the same application representing a shared design identity~\cite{chai2025amex}. To adapt the dataset for studying design conformance, we first removed screenshots corresponding to system UIs, such as home screens and app launchers. We then used an off-the-shelf model~\cite{wu2024uiclip} to remove highly similar screens. After filtering and de-duplication, the dataset contained 2,071 screenshots from 52 app packages. We split these packages into 40 training, 6 validation, and 6 test packages.

\subsubsection{Conformance Relationships}

We operationalized our conceptualization of conformance by treating existing app screens as partial evidence of a designer's latent judgment \(c(y)\). For each design context, a reference set \(R\) contained screens from the same app package that are known to share a design identity. Candidate pairs \((x^+,x^-)\) represent designs whose conformance to that identity must be compared. In an interactive setting, a designer could directly indicate which candidate better matches their intent. However, for offline analysis, we constructed pairwise orderings from relationships already present in the dataset and from controlled synthetic variations, with \(x^+\) expected to conform more closely to \(R\) than \(x^-\).

We construct three types of conformance relationships. Package identity compares designs across different apps, screen provenance compares an original design with its reconstruction, and interpolation creates controlled variants between two design identities. Our data generation approach is depicted in Figure \ref{fig:data-generation}.

\paragraph{Package Identity.}
Our first relationship comes directly from the structure of the AMEX dataset. Screens from the same app package share a design identity, while screens from different packages provide contrasting examples. Given a reference set \(R\), we treat a held-out screen \(x^+\) from the same package as more conformant than a screen \(x^-\) drawn from another package. These pairs test whether designs associated with different design identities can be distinguished from one another.

\paragraph{Screen Provenance.}While package identity provided only coarse comparisons between design systems, it does not show how conformance varies within the same underlying interface. To create finer-grained comparisons, we reconstructed selected AMEX screens as editable HTML/CSS representations using a multimodal language model~\footnote{Qwen/Qwen3.5-35B-A3B}.

The reconstruction process provided more controlled comparison in which the underlying content and structure remain largely fixed. We treated the original AMEX screenshot \(x^+\) as more conformant to references from its package than its reconstructed counterpart \(x^-\), which likely introduced small deviations in properties such as color, spacing, or component appearance.

\paragraph{Interpolated Designs.}
To create graded differences in conformance, we gradually altered a reconstructed screen progressively toward the design identity of a foreign donor package, following prior work on design interpolation~\cite{lu2025misty,wu2024framekit}. We first adapted the screen to the donor design, then diff the original and adapted HTML/CSS and apply 25\%, 50\%, 75\%, or 100\% of the resulting code patches. This produced a sequence of variants spanning the source and donor identities, which we rendered into new screens. Relative to the original references \(R\), we expected variants with less donor styling to rank higher in conformance. We expected the opposite conformance label relative to donor references \(D\). We generated an average of 33.8 variations for every real image, where repetitions were generated against randomly sampled foreign donor packages.

\begin{figure*}[t]
    \centering
    \includegraphics[width=\textwidth]{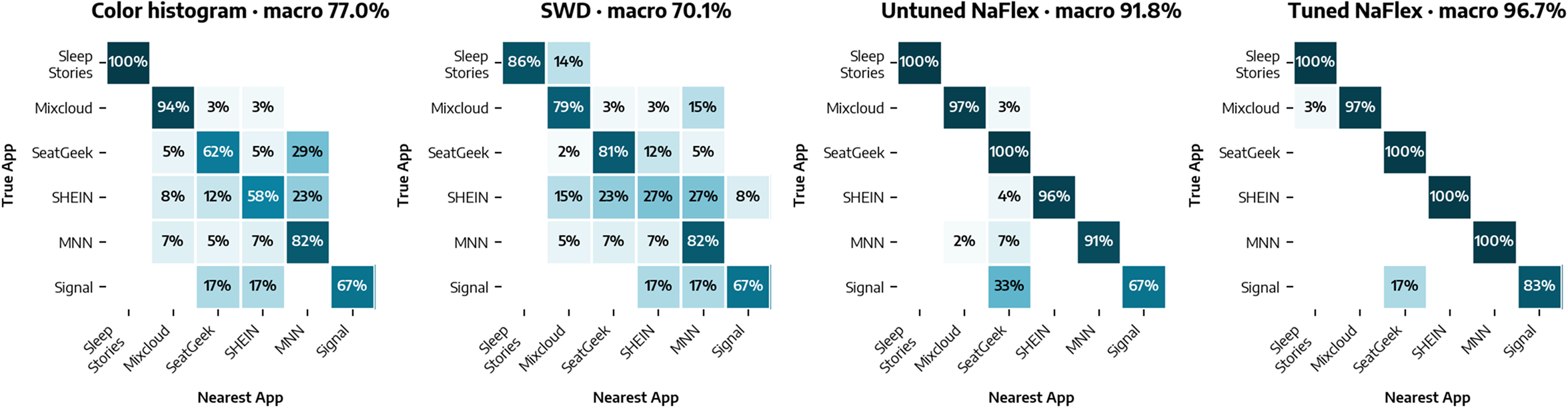}
    \caption{Nearest-neighbor accuracy by app package (rows) and predicted nearest app (columns), compared across four similarity methods (color histogram, SWD, untuned SigLIP2, tuned model).}
    \Description{This image displays four side-by-side confusion matrices evaluating different models on identifying the correct app package for a given screenshot. The y-axis represents the "True App," and the x-axis represents the "Nearest App," featuring six specific apps: Sleep Stories, Mixcloud, SeatGeek, SHEIN, MNN, and Signal. The matrices evaluate, from left to right, "Color histogram" (macro 77.0\%), "SWD" (macro 70.1\%), "Untuned NaFlex" (macro 91.8\%), and "Tuned NaFlex" (macro 96.7\%). Using a blue color scale where darker shades denote higher percentages, the visual progression clearly demonstrates that the "Tuned NaFlex" model performs the best, indicated by the darkest and highest-percentage diagonal line representing correct matches.}
    \label{fig:test-confusion}
\end{figure*}
\subsubsection{Model Architecture}
\label{sec:model-architecture}
\begin{figure}[!htb]
    \centering
    \includegraphics[width=\columnwidth]{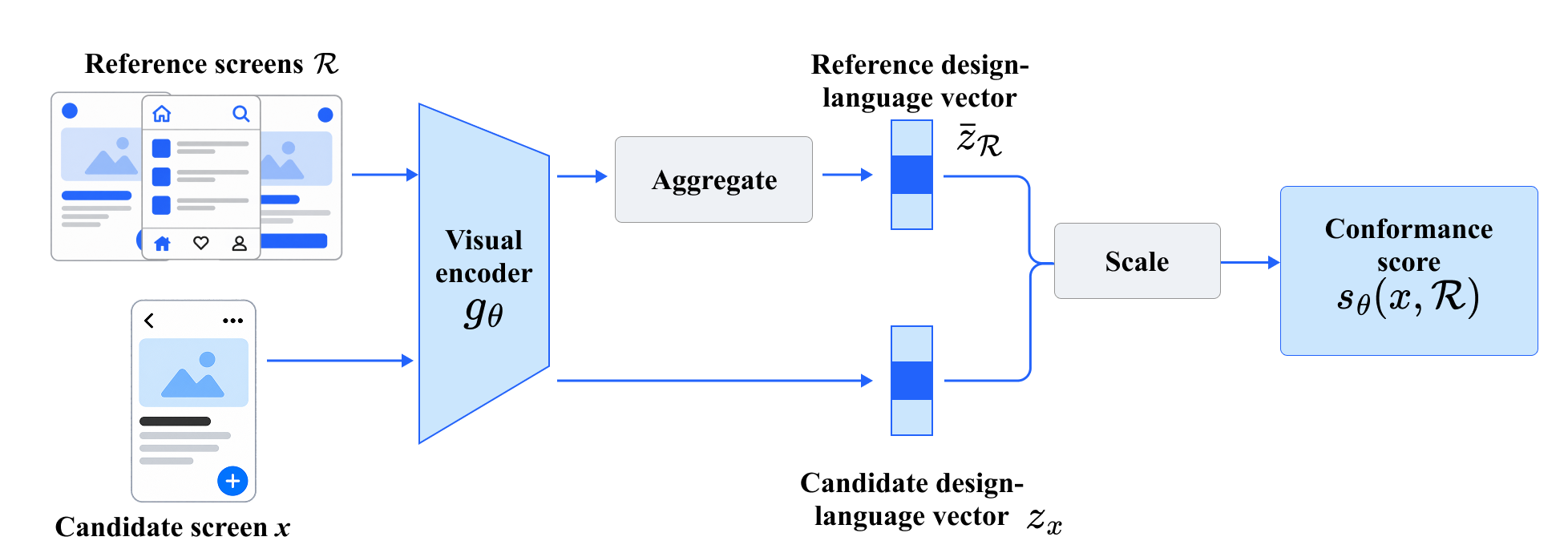}
    \caption{The conformance scoring pipeline, showing how reference and candidate screens are visually encoded, aggregated, and compared to yield a final score.}
    \Description{This flowchart illustrates the conformance scoring pipeline. It begins with "Reference screens $\mathcal{R}$" and a "Candidate screen $x$" both being processed by a "Visual encoder $g_\theta$". The encoded reference screens are then passed through an "Aggregate" step to produce a "Reference design-language vector $\bar{z}_\mathcal{R}$", while the encoded candidate screen directly outputs a "Candidate design-language vector $z_x$". Finally, both vectors are fed together into a "Scale" operation, which computes the final "Conformance score $s_\theta(x, \mathcal{R})$".}
    \label{fig:model-architecture}
\end{figure}
Our model architecture (Figure \ref{fig:model-architecture}) estimates design conformance of a candidate by comparing it against a representation derived from set of reference designs and additional designer feedback in a learned embedding space.
Following our conceptual formulation, the references provide partial evidence as instances sampled from the true ``region'' of designs that a designer might consider highly conformant $c(y)$.
Using multiple references, our model estimates how the designers' tacit conformant region is mapped to its learned embedding space then uses a similarity metric (e.g., cosine similarity) to evaluate a candidate based on its proximity to this region.

Our model first uses a visual encoder $g(x)$ that generates an embedding $z_x$ from an input screenshot $x$.
We treat $z_x$ as a representation of the visual design style of $x$, rather than its  content, purpose~\cite{wang2021screen2words}, or overall design quality~\cite{wu2024uiclip,wu2026improving}.
Following prior work~\cite{wu2024uiclip,wu2026improving}, we initialize \(g\) from an off-the-shelf CLIP model to leverage its large-scale pretraining. Specifically, we use a SigLIP2 NaFlex, which is more recent variant of the CLIP model architecture which incorporated more data in its pre-training phase and supports non-square image aspect ratios (common in UI designs)~\cite{tschannen2025siglip2}.

Because individual reference embeddings are partial and noisy estimates of their underlying design identity, we initialize the reference prototype $z_R$ using their normalized mean.

\begin{equation}
    \bar{z}_R =
    \operatorname{normalize}
    \left(
        \frac{1}{|R|}
        \sum_{r \in R} g(r)
    \right).
\end{equation}

We explored two ways that this initial estimate can be further updated by designer interactions. First, as designers add references, \(z_R\) improves by averaging over a larger set of observed examples and shifts the mean prototype towards a location that better describes all available partial evidence. If the preference information is also available (i.e., design A is more conformant than design B), a more targeted update vector can be applied to the prototype by i) computing the average embeddings of ``rejected'' designs, ii) subtracting this from the embedding of the ``chosen design'', and iii) projecting this difference onto the subspace orthogonal to the initial reference prototype, then adding it to the prototype and renormalizing it.

We score a candidate \(x\) by computing a linearly scaled cosine similarity measure between its embedding $z_x$ and the estimated prototype $z_R$.

\begin{equation}
    s(x,R)
    =
    \alpha \cdot
    \cos\left(g(x),\bar{z}_R\right)-c.
\end{equation}

Because SigLIP2’s pretrained similarity scale is not centered at zero, we learn \(\alpha\) and \(c\) to rescale and recenter the cosine scores, which substantially improved training efficiency.

\subsubsection{Modeling Experiments}
\begin{figure*}[t]
    \centering
    \includegraphics[width=0.32\textwidth]{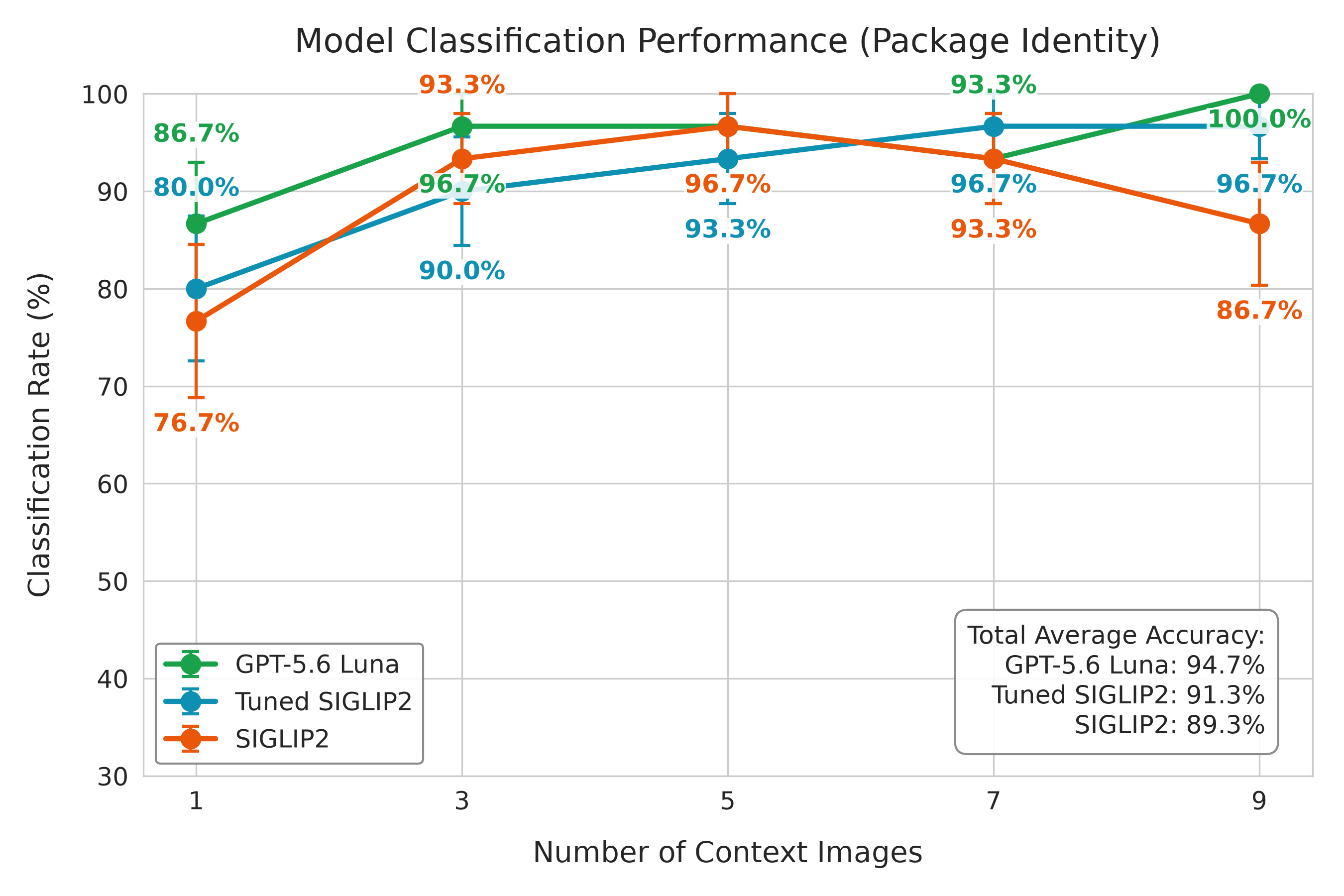}
    \hfill
    \includegraphics[width=0.32\textwidth]{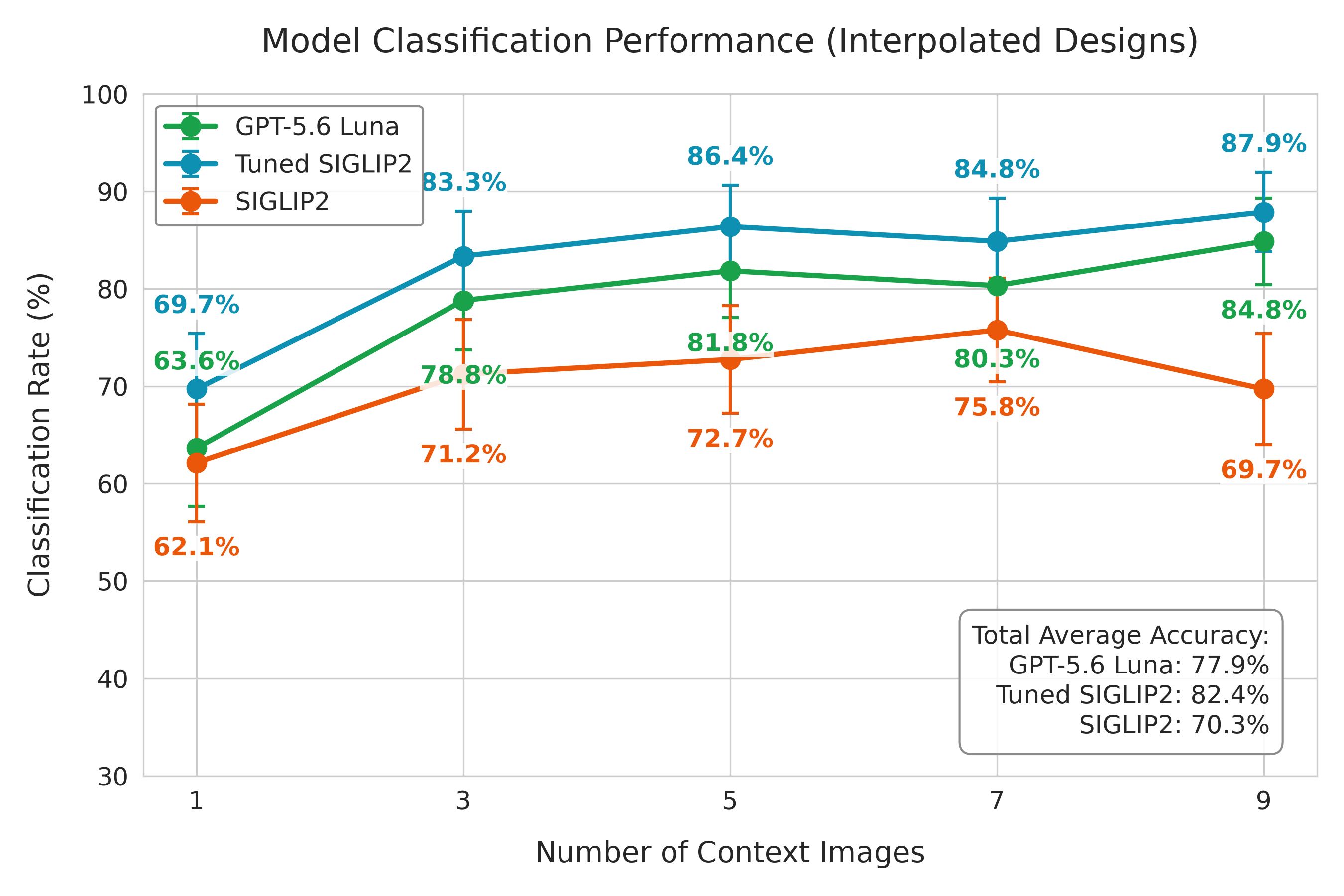}
    \includegraphics[width=0.32\textwidth]{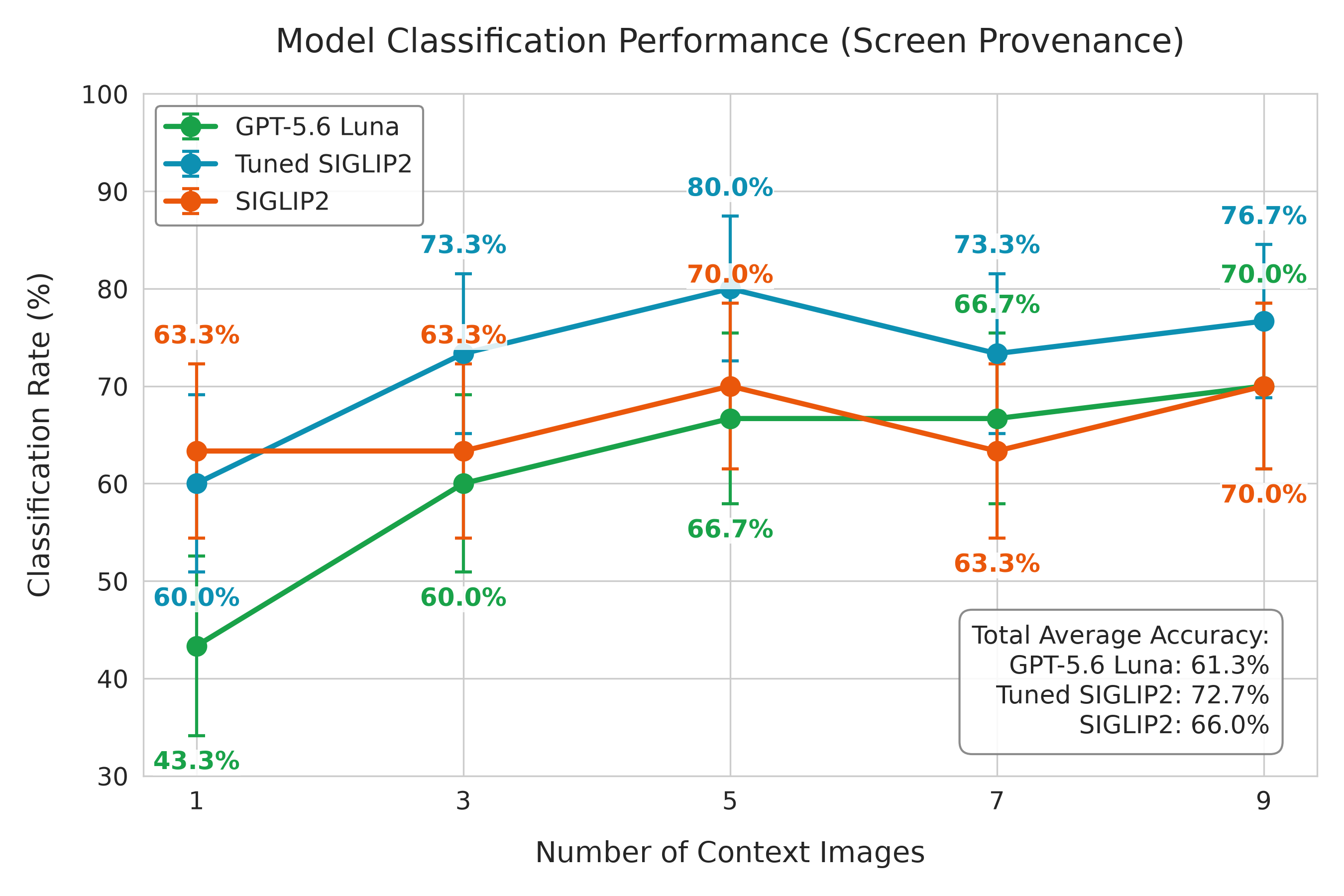}
    \caption{Conformance classification accuracy across the number of reference images, by relation type (package identity, screen provenance, interpolated designs) and model (GPT-5.6 Luna, tuned SigLIP2, untuned SigLIP2).}
    \Description{This figure features three side-by-side line graphs demonstrating conformance classification accuracy across an increasing number of context images (1 through 9 on the x-axis). Each graph evaluates three models—GPT-5.6 Luna (green line), Tuned SigLIP2 (blue line), and untuned SigLIP2 (orange line)—across different relation types. The left graph, "Package Identity," shows all models performing highly, with GPT-5.6 Luna achieving the highest total average accuracy of 94.7\%. In the middle graph, "Interpolated Designs," and the right graph, "Screen Provenance," Tuned SigLIP2 consistently outperforms the other models, reaching the highest total average accuracies of 82.4\% and 72.7\%, respectively. Across all three charts, classification rates generally trend upward as the number of context reference images increases.}
    \label{fig:model-validation} %
\end{figure*}

We trained a generalized conformance model on our dataset and conducted two technical evaluations to assess i) how well different approaches represent design-system identity and ii) how well different machine learning models can predict conformance relationships.

We initialized the visual encoder from the publicly released SigLIP2 Base NaFlex checkpoint and trained it on a mix of the pairwise comparisons from the 40 training application packages.
During training, we sampled a reference set \(R\) corresponding to a real UI screen.
Then we randomly sampled one type of conformance relation for each relation type corresponding to that real anchor (e.g., package identity), resulting in multiple pairs \((x^+,x^-)\).
We noticed that for some relationships, such as screen provenance, the presence of the system header bar and navigation UI from real UI screenshots made it much easier for the model to make correct predictions without actually understanding the screen's design.
As a precaution, we applied a preprocessing pipeline where we cropped off 10\% from the top and bottom of all screenshots during training and evaluation of AMEX-derived data.
We optimized the model using a pairwise contrastive objective that trained the model to score \(x^+\) above \(x^-\) by a certain margin $m=0.5$ for each training pair, where $m$ was chosen through hyperparameter tuning experiments. 
\begin{equation}
    \mathcal{L}_{\mathrm{rank}}
    =
    \max\left(
        0,
        m - s(x^+,R) + s(x^-,R)
    \right),
\end{equation}

We combined gradients from the package identity, screen provenance, and interpolated-design comparisons using a technique called PCGrad~\cite{yu2020gradient}, which is used in multi-task learning for computing more efficient, non-conflicting updates. We varied the size of \(R\) by selecting a random subset of reference screenshots available in an application to improve cardinality generalization and updated only the final vision-transformer layer and scoring parameters to reduce the chance of overfitting.
We trained for a maximum of 4000 optimizer steps, where we used a batch size consisting of examples from 8 real anchors at a time.
We selected the final checkpoint using the best loss on six validation packages, which occurred at 3750 steps.
Additional hyperparameters are provided in the appendix.

\paragraph{Design System Representation.}

We first studied how well different metrics distinguish visual variation within a design system from variation across design systems.
To do this, we conducted a nearest-neighbor lookup protocol for screens inside of our test set.
For each screenshot in the held-out test set, we retrieved the nearest other screenshot according to three commonly-used baseline metrics and our tuned model and measured whether the nearest neighbor came from the same app package.
To serve as a realistic set of comparison points, we chose to use i) color histogram distance, ii) Sliced Wasserstein Distance (SWD)~\cite{karras2018progressive}, iii) an untuned version of the SigLIP2 model, and iv) our tuned model.
At a high level, i) can capture global color frequencies, ii) measure the similarity of two images by comparing patches at different scales, and iii) represent an off-the-shelf neural embedding method trained for image-text semantics.

The results of our experiments are shown in Figure~\ref{fig:test-confusion}, the tuned encoder achieved the highest macro accuracy at 96.7\%, compared with 91.8\% for untuned NaFlex, 77.0\% for color histograms, and 70.1\% for SWD.
First, our results suggest that design system identity cannot be easily captured by off-the-shelf metrics that focus on a single aspect of design, e.g., color or structural patch similarity.
Each of these metrics alone might cause different types of apps to be confused with each other, evident in the classification errors of the color histogram and SWD metric.
Our tuned model performed the best and had considerable improvement over the pre-trained baseline, which could also theoretically incorporate different types of color and layout information, suggesting that our dataset was useful for this type of representation learning.
\paragraph{Design Conformance Prediction.}
In addition to representation quality, we conducted an experiment to evaluate performance on predicting known design conformance relations.

In this experiment, our main point of comparison was the untuned SigLIP2 NaFlex model and GPT-5.6 Luna, a proprietary cloud LLM~\cite{openai2026gpt56luna}.
Conformance queries were formulated to the LLM using a prompt template that included a set of reference images and asked the model to select the more conformant design between two candidates (see Appendix).

We evaluated each model on conformance relations constructed from images in our test set, also applying preprocessing steps that cropped out system header and navigation elements.
We varied the number of reference images to test how each model used different amounts of evidence about the design system. For each candidate pair, we evaluated \(N \in \{1,3,5,7,9\}\) randomly selected reference images (excluding the candidate's anchor). If a test example's application did not contain enough reference images, the model used $\min(N, |R|)$.

The results of our experiment are shown in Figure \ref{fig:model-validation}.
In general, there a large effect on the type of conformance relation and each models' performance, suggesting that some are ``easier'' to detect that others.
Although metrics based on color distributions and image structure can struggle to differentiate between app designs, the machine learning models used in this experiment had strong performance in predicting package identity relations.
GPT-5.6 Luna achieved the highest performance in detecting package identities (94.7\%), followed closely by our tuned model (91.3\%).
On the other hand, screen provenance and design interpolation relations had lower performance across all tested conditions, and our tuned model achieved the best prediction accuracy among all models for both, 82.4\% and 72.7\%, respectively. In general, the models achieved better performance when more references were used, with the largest improvement occurring between 1 and 3 references.

\begin{tcolorbox}[
    colback=guidelightblue!35!white,
    colframe=guideblue!20!white,
    boxrule=0.4pt,
    left=6pt,
    right=6pt,
    top=4pt,
    bottom=4pt,
    arc=1pt
]
Our experiments showed that design identity, a target of conformance, depends on interactions among visual properties rather than any single property such as color or regional comparisons. These relationships can be learned from data to represent design systems and predict relative conformance.
\end{tcolorbox}

\section{GUIDE}
In this section, we introduce GUIDE, a designer-in-the-loop authoring pipeline for conformant GenUIs. GUIDE is a Figma plugin that allows designers to align a GenUI. Internally, the GenUI uses learnable prompts, examples, and the previously-introduced conformance scoring model (Figure \ref{fig:system-pipeline}) to generate conformant designs from a screen description. As designers interact with the GenUI through the plugin, GUIDE updates the GenUI's internals, refining its behavior to match the designers' intention. 

\subsection{Walkthrough}
\begin{figure*}[!htb]
    \centering
    \includegraphics[width=\textwidth]{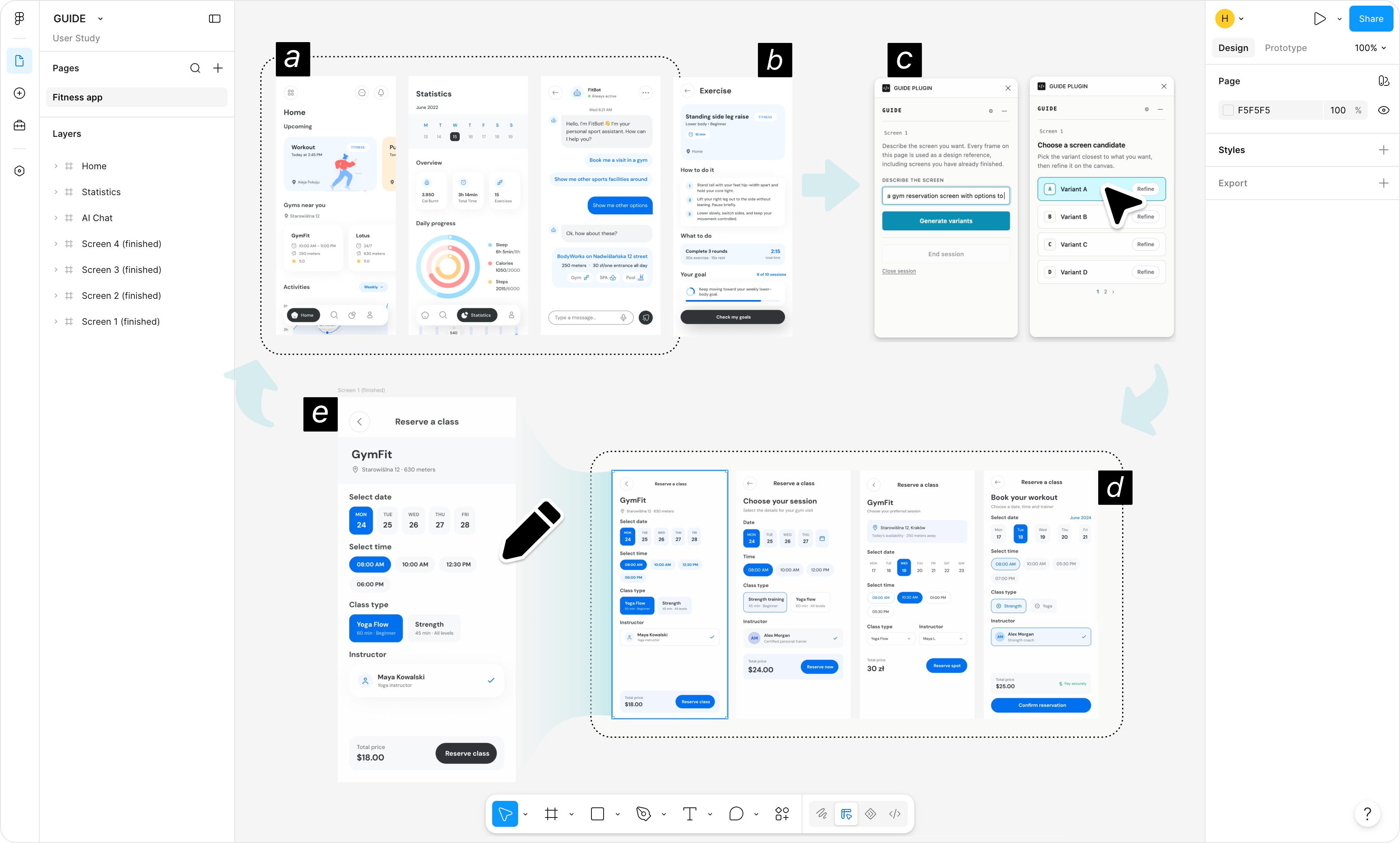}
    \caption{Walkthrough example of a designer extending \textit{Fitness Tracking App} with GUIDE. (a) The app's three original screens. (b) The \textit{Exercise Instruction} screen the designer added in the previous iteration. (c) The designer prompts a \textit{Gym Reservation} screen and (d) selects one of the generated candidates. (e) The designer refines the screen to be more conformant with the app.}
    \Description{This screenshot displays a design interface illustrating a walkthrough of a designer extending a "Fitness Tracking App" using the GUIDE tool. The process is broken into five steps connected by arrows in a cycle: steps (a) and (b) show the app's three original screens next to an "Exercise Instruction" screen added in a previous iteration; step (c) shows the designer using a plugin panel to prompt for a "Gym Reservation" screen; step (d) displays a panel of generated screen candidates where the designer selects one; and step (e) highlights the chosen candidate being manually edited (indicated by a pen icon) to improve its visual conformance to the rest of the app before looping back to the start.}
    \label{fig:interface}
\end{figure*}

In this section, we present a walkthrough to illustrate how Alex, a designer, would use GUIDE to align a GenUI to the design system of an existing artboard, \textit{Fitness Tracking App}\footnote{Concept by Miquido and Filip Rygucki, licensed under CC BY 4.0:
\url{https://www.figma.com/community/file/1124291019178940578}.} (Figure~\ref{fig:interface}).

\subsubsection{Inititalization}

Alex first opens the artboard corresponding to the design system they wish for the GenUI to follow.
GUIDE incorporates designs on the artboard (a) into i) the GenUI's prompt as references and instructions and ii) signal for the GenUI's internal scoring model.
Through sustained use, this body of references will grow to include the existing app screens (a) and new examples created through the alignment process, such as the \textit{Exercise Instruction} screen (b).

\subsubsection{Generating Candidates}

To inspect the design that the GenUI would create for a \textit{Gym Reservation} screen, Alex invokes the plugin window which allows them to query GenUI for a given prompt.
Alex authors a more specific prompt describing the screen they want to create (c) focusing on the purpose of the screen.
GUIDE calls the GenUI to generate several screen candidates, which are inserted as editable objects on the canvas (d). Different candidates might reflect different ways of satisfying the prompt given its encoded design constraints.
Alex reviews and selects the candidate closest to their imagined design before using the selection as the starting point for manual refinement. From a practical point of view, selecting a generated, mostly-conformant candidate as the starting point allows them to reduce effort needed to create additional references. From a conceptual point of view, the act of examining the variants also gives the designer insight into the GenUI's interpretation of their intent and true output distribution.

\subsubsection{Refining and Accepting a Candidate}

To refine a design, Alex can use familiar Figma interactions, e.g., changing properties or replacing components, to edit and modify their selected candidate (an ordinary Figma frame) (e).

When Alex is satisfied with their design, they mark the frame as ``complete'' in the plugin, which will update the GenUI's internal prompt and scoring models in several ways.
We describe GUIDE's pipeline in more detail in the following section, but this process is mostly invisible to Alex who can continue to align the GenUI through repeated inspection, authoring, and refinement.
Finally, when Alex is satisfied with the GenUI's performance, it can be converted into a portable package of resources, prompts, and weights that can be used to run the GenUI outside of the plugin.

\begin{figure}[t]
    \centering
    \includegraphics[width=\columnwidth]{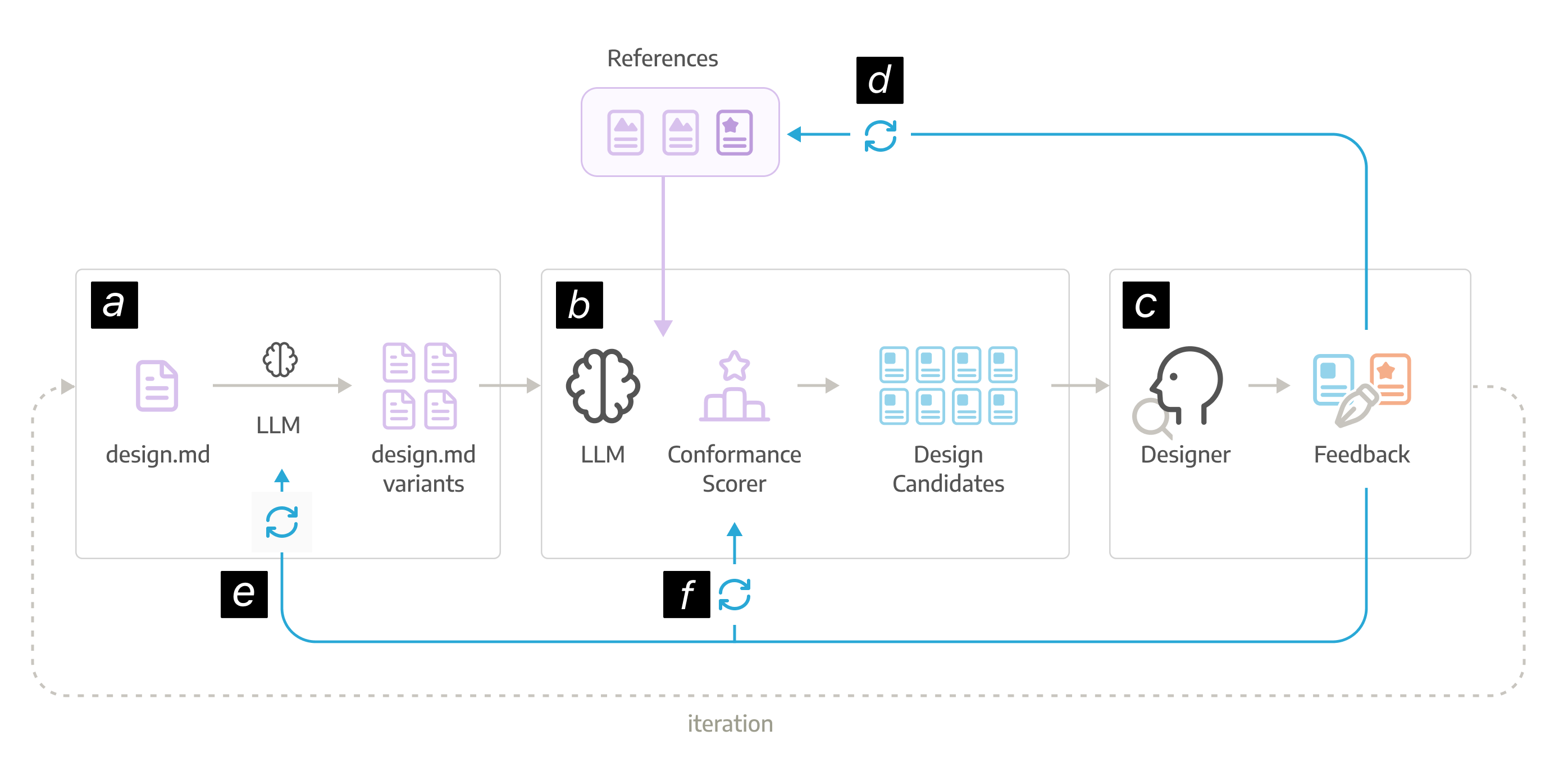}
    \caption{The GUIDE pipeline, illustrating how the system works with the \textit{GenUI system} based on designer interaction. (a) \texttt{design.md} synthesis, (b) screen generation and scoring, (c) collecting designer feedback and updating each component of the \textit{GenUI system}: (d) reference artboards, (e) \texttt{design.md}, and (f) the conformance scorer.}
    \label{fig:system-pipeline}
    \Description{This flowchart illustrates the GUIDE pipeline's iterative workflow across three main stages. In stage (a), an initial \texttt{design.md} file is processed by an LLM to synthesize \texttt{design.md} variants. In stage (b), these variants, combined with a set of "References," are used by an LLM and a "Conformance Scorer" to produce a set of "Design Candidates." In stage (c), a "Designer" reviews these candidates to provide "Feedback." This feedback loops back via three distinct blue update paths to align the GenUI system: path (d) updates the References, path (e) updates the original \texttt{design.md} file, and path (f) updates the Conformance Scorer. The main stages (a, b, and c) are enclosed within a dotted boundary labeled "Iteration."}
\end{figure}

\subsection{Pipeline}
\label{sec:pipeline}
Figure \ref{fig:system-pipeline} summarizes the pipeline that GUIDE uses to update a GenUI system from designer interaction. In this section, we describe each component in more detail.

\subsubsection{(a) \texttt{design.md} Synthesis}
GUIDE aligns the GenUI in part through the generator's input, which contains a fixed prompt format (consisting of general instructions), a textual design document (\texttt{design.md}\footnote{https://github.com/google-labs-code/design.md}) learned from the project, and reference exemplars.
The \texttt{design.md} document is synthesized during initialization, i.e., when GUIDE is first enabled on a project.
GUIDE uses a prompt to traverse and process content on the artboard to build a collection of tokens (e.g., colors, typography, components, and effects) from on-canvas designs. These tokens are fed into another prompt that uses extracted images of the given references generate four potential \texttt{design.md} files (through repeated sampling) that encapsulate the design language of the project.
The next round of designer feedback determines which of the four candidates is retained.

\subsubsection{(b) Screen Generation and Scoring}
When the designer uses GUIDE to query the GenUI with a screen description, eight candidates are generated. The same set of reference screens and scoring model prototype is used across all eight screens, while each of the four \texttt{design.md} variants is used twice. The LLM is instructed to generate UI output as an HTML fragment with a limited set of inline CSS rules that are supported by our HTML to Figma importer~\footnote{https://github.com/vercel/satori} (see Appendix). The plugin also maintains PNG renderings of all on-canvas references, which are the expected input format of the scoring model.
During the GenUI design phase, the scoring model is not activated, but during deployment, the GenUI uses best-of-n sampling by automatically selecting the top scoring variant among eight candidates.

\subsubsection{Updating the references (d), prompt (e), and scorer (f)}
After the eight generated candidates are converted into Figma, the designer selects one and edits it to address any remaining issues (c). This triggers several updates to the GenUI system.

\begin{itemize}
    \item The \texttt{design.md} associated with that candidate becomes the new base \texttt{design.md}, while the finalized screen is added to the reference set (d).
    \item A new set of four candidate \texttt{design.md} prompts is generated for the designer to choose from in future rounds of feedback (e). GUIDE uses a prompt with the before-and-after screens and structured JSON representation of the changed properties and generates four updated \texttt{design.md} candidates through temperature-based resampling. Because the intent behind an edit can be ambiguous, these alternatives capture different possible interpretations of how the \texttt{design.md} should be updated.
    \item The scoring model is updated by recomputing its prototype vector $\bar{z}_R$ based on additionally observed evidence. We explored both update methods described in Section \ref{sec:model-architecture}, which included updating $\bar{z}_R$ based on re-computing the average of the reference set, and computing a directional vector using the difference between the chosen and rejected variants.
\end{itemize}

\begin{tcolorbox}[
    colback=guidelightblue!35!white,
    colframe=guideblue!20!white,
    boxrule=0.4pt,
    left=6pt,
    right=6pt,
    top=4pt,
    bottom=4pt,
    arc=1pt
]
The output of GenUIs can be controlled by editing its i) prompt, ii) in-context examples, and iii) inference technique. GUIDE is a system that allows designers to change these artifacts (i-iii) in real-time using a pipeline that converts natural designer interactions into artifact-specific updates.
\end{tcolorbox}

\section{User Study}
To evaluate the usability of GUIDE and its performance in aligning GenUIs, we conducted a user study with 12 UI/UX practitioners. 
\subsection{Methodology}
\subsubsection{Participants}
We recruited 12 UI/UX practitioners from the authors' professional network. All participants reported having prior experience in UI/UX, either through pursuing a design-related degree or through design project experience. Their years of experience ranged from 1.5 to 6 years (M = 3.67, SD = 1.21), and their self-reported proficiency (scale  of 1–10) ranged from 5 to 8 (M = 6.83, SD = 1.11). Participants were compensated with a \$50 gift card for a 2-hour study session.

\subsubsection{Procedure}
All sessions were conducted via Zoom and recorded with participants' consent, and lasted 120 minutes in total. 
This study was approved by our institution's IRB.

\paragraph{Tutorial (15 minutes).} With a brief introduction, participants installed the plugin, followed the tutorial, practiced shortcuts, and completed a practice task using the tutorial artboards.
\paragraph{Main Session (90 minutes).} They then completed 2 sessions, each consisting of 4 screen generation and modification tasks (35 minutes) followed by a model evaluation ``arena''~\cite{chiang2024chatbot} (10 minutes).
At the start of each session, participants had 1 minute to review the existing reference artboards (e.g., the home screen and dashboard screen of a fitness app). For each screen task, participants were given a screen keyword to design (e.g., "profile screen" of a fitness app) and asked to ideate accordingly. They then wrote a screen description reflecting their design idea (e.g., "a screen where users can check their profile and edit contents including name, location, and workout preference"), optionally including specific details such as layout or style. Participants reviewed 8 design candidates generated by the plugin and, within 6 minutes, selected and modified one based on their idea and their own criteria for conformance. This process was repeated so that each participant authored 4 new screens per artboard.
At the end of each session, participants completed a 10-minute evaluation arena. This arena consisted of 10 pairs of screens corresponding to a held-out set of screen descriptions. Each pair contained one screen generated by the GenUI system before any of their feedback, i.e., initial artboard references, generated \texttt{design.md}, and scoring model prototype, and one screen generated by their final GenUI system. For each pair, they were asked to select the candidate that they felt was more conformant with the corresponding artboard's design identity.
\paragraph{Interview (15 minutes).} After completing both sessions (i.e., artboards), a brief interview was conducted, focusing on the overall experience of the participants and their perception of conformance. See Appendix for the full list of questions.

\subsubsection{Study Materials}
To reduce the effect of the given reference styles on model performance, we constructed 2 sets of study materials, each composed of 2 reference artboard sets, with each set containing 2–3 screens.
We intentionally limited artboards in each set to 2–3 references so that the design language remained only partially specified, leaving room for participants to align the GenUI according to their own interpretation.

\begin{itemize}
    \item \textbf{Set 1}
    \begin{itemize}
        \item Session 1: Learning Management App (2 screens)
        \item Session 2: Fitness Tracking App (3 screens)
    \end{itemize}
    \item \textbf{Set 2}
    \begin{itemize}
        \item Session 1: Food Delivery App (2 screens)
        \item Session 2: Digital Payments App (3 screens)
    \end{itemize}
\end{itemize}

Each half of the study was conducted with one of the two materials, which were randomly assigned to participants. All reference artboards were collected from the Figma community and were publicly shared under the Creative Commons Attribution 4.0 (CC BY 4.0) license, all of which are mobile apps.

We provided four broad screen concepts (e.g., ``Instructor Profile'') that did not overlap with the provided screens but could plausibly belong to the same app. Each concept left room for participants to interpret the screen through a more detailed prompt. We chose our screen concepts to hint at screens that might exist at different levels of the app's information architecture, e.g., top-level and sub-level screens. The 10 descriptions used for the evaluation arena of each artboard were then constructed by expanding 2 screen concepts that were also used by participants, 3 screen concepts similar to ones used by participants, and 5 screen concepts that were not used at all. We provide the artboards, screen concepts, and screen prompts in the Appendix of this paper.


\subsection{Quantitative Results}

In this section, we report quantitative data collected from the user study, covering i) how participants interacted with GUIDE, ii) how they evaluated the aligned GenUI, and iii) how the GenUI's alignment developed over successive rounds of feedback.

\begin{figure}[t]
    \centering
    \includegraphics[width=\columnwidth]{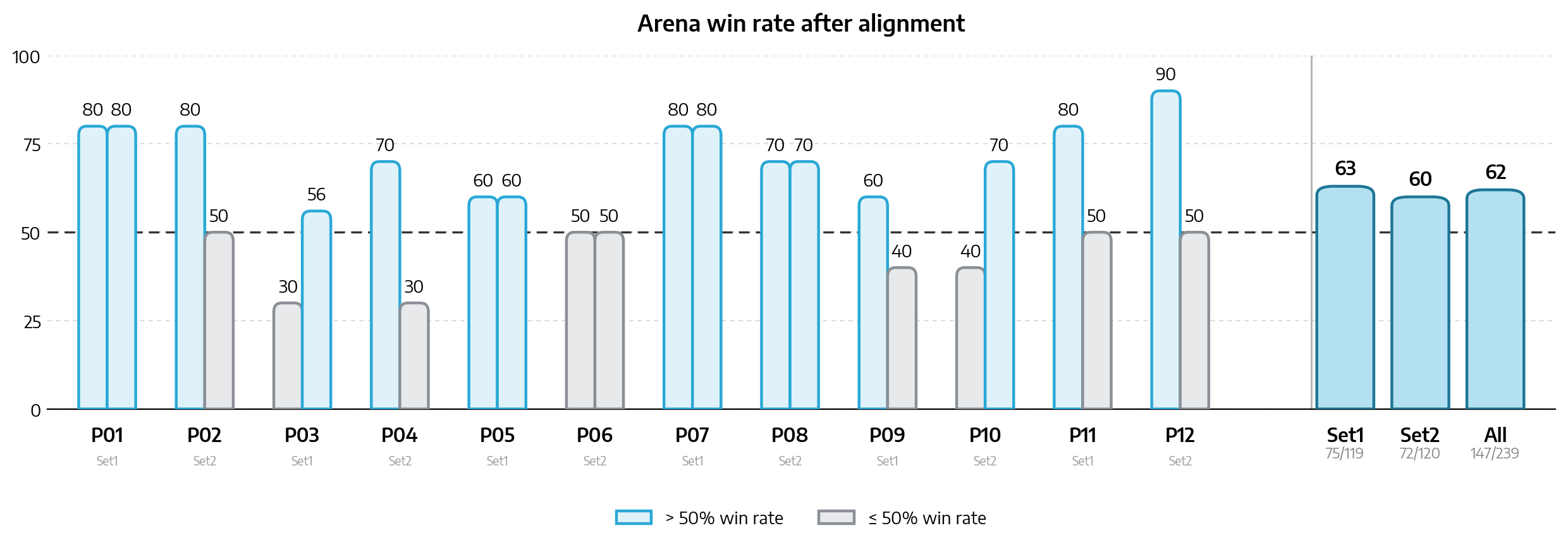}
    \caption{Arena win rates for the final-aligned GenUI system across 12 participants, each assigned one of 2 material sets, achieving an overall win rate of 62\%.}
    \Description{This bar chart visualizes the "Arena win rate after alignment" across 12 individual participants (P01-P12) alongside aggregate scores. Each participant displays two bars evaluated against a 50\% threshold, which is marked by a dashed horizontal line. Scores greater than 50\% are colored blue, while scores at or below 50\% are grey. The majority of the individual bars are blue, with participant P12 reaching the highest single score of 90\%. A separate section on the right displays the aggregate performance, showing that Set 1 achieved a 63\% win rate and Set 2 achieved 60\%, culminating in an overall combined win rate of 62\% across all trials.}
    \label{fig:arena-results}
\end{figure}

\begin{figure}[t]
    \centering

    \begin{minipage}[t]{0.49\linewidth}
        \centering
        \includegraphics[width=\linewidth]{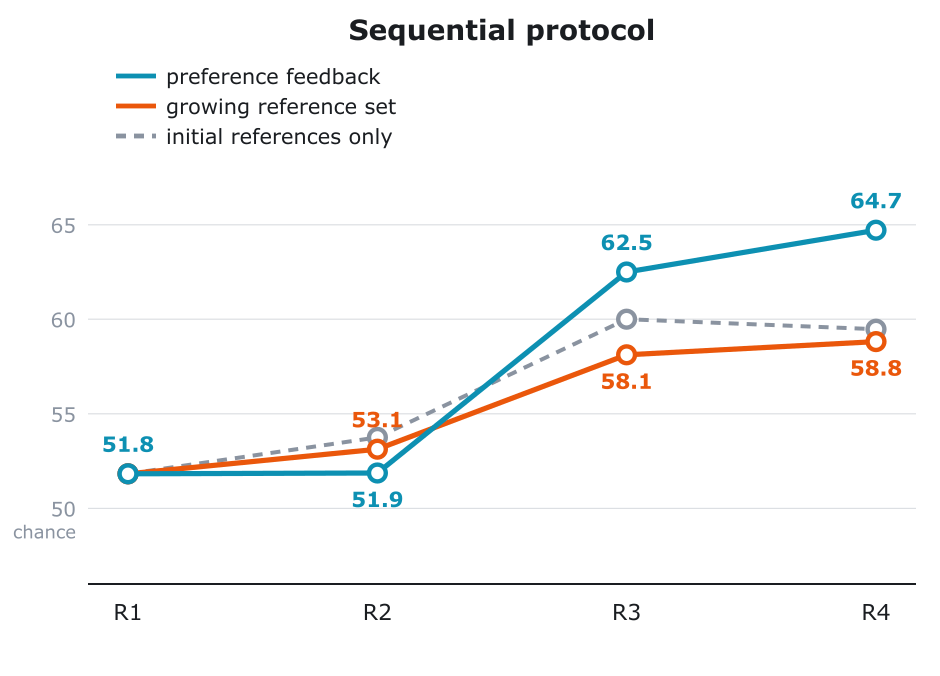}
    \end{minipage}
    \hfill
    \begin{minipage}[t]{0.49\linewidth}
        \centering
        \includegraphics[width=\linewidth]{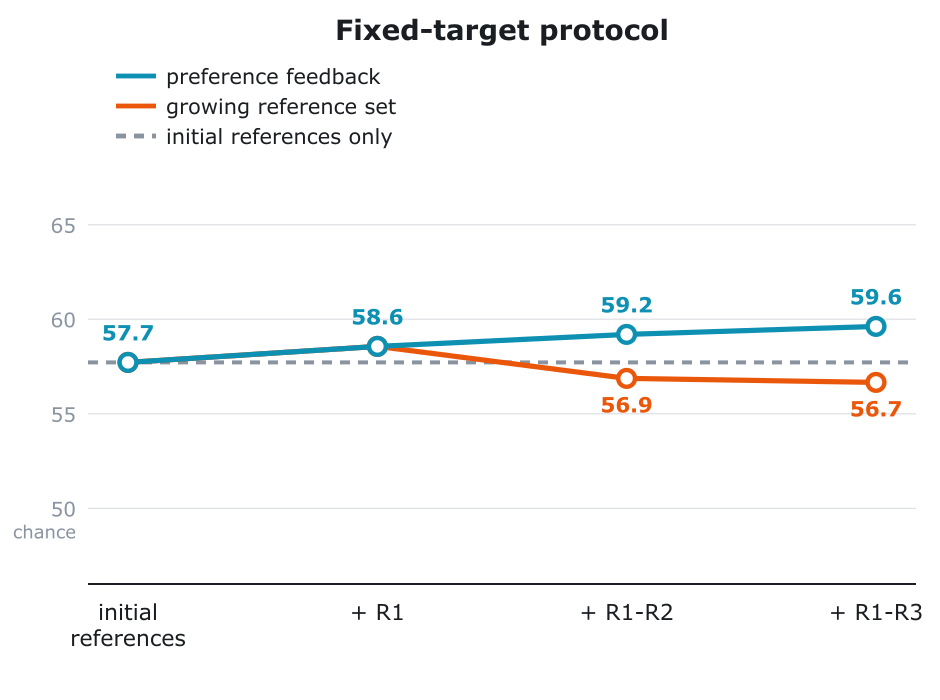}
    \end{minipage}

    \caption{Win rate (\%) across alignment rounds (R1-R4), comparing preference feedback, growing reference set, and initial-references-only strategies, under sequential and fixed-target protocols.}
    \Description{This image features two side-by-side line graphs comparing performance across multiple alignment rounds using three strategies: "preference feedback" (solid blue line), "growing reference set" (solid orange line), and "initial references only" (grey dashed line). The left graph, "Sequential protocol," shows all strategies starting near 51.8 at R1 and generally increasing over time, with preference feedback reaching the highest peak of 64.7 at R4. The right graph, "Fixed-target protocol," starts at a baseline of 57.7; as rounds progress, preference feedback steadily climbs to a high of 59.6, while the growing reference set slowly declines to 56.7. The accompanying caption concludes that preference feedback consistently yields higher win rates than the alternative reference strategies in both tested protocols.}
    \label{fig:overall-scorer}
\end{figure}

\begin{figure*}[t]
    \centering
    \includegraphics[width=\textwidth]{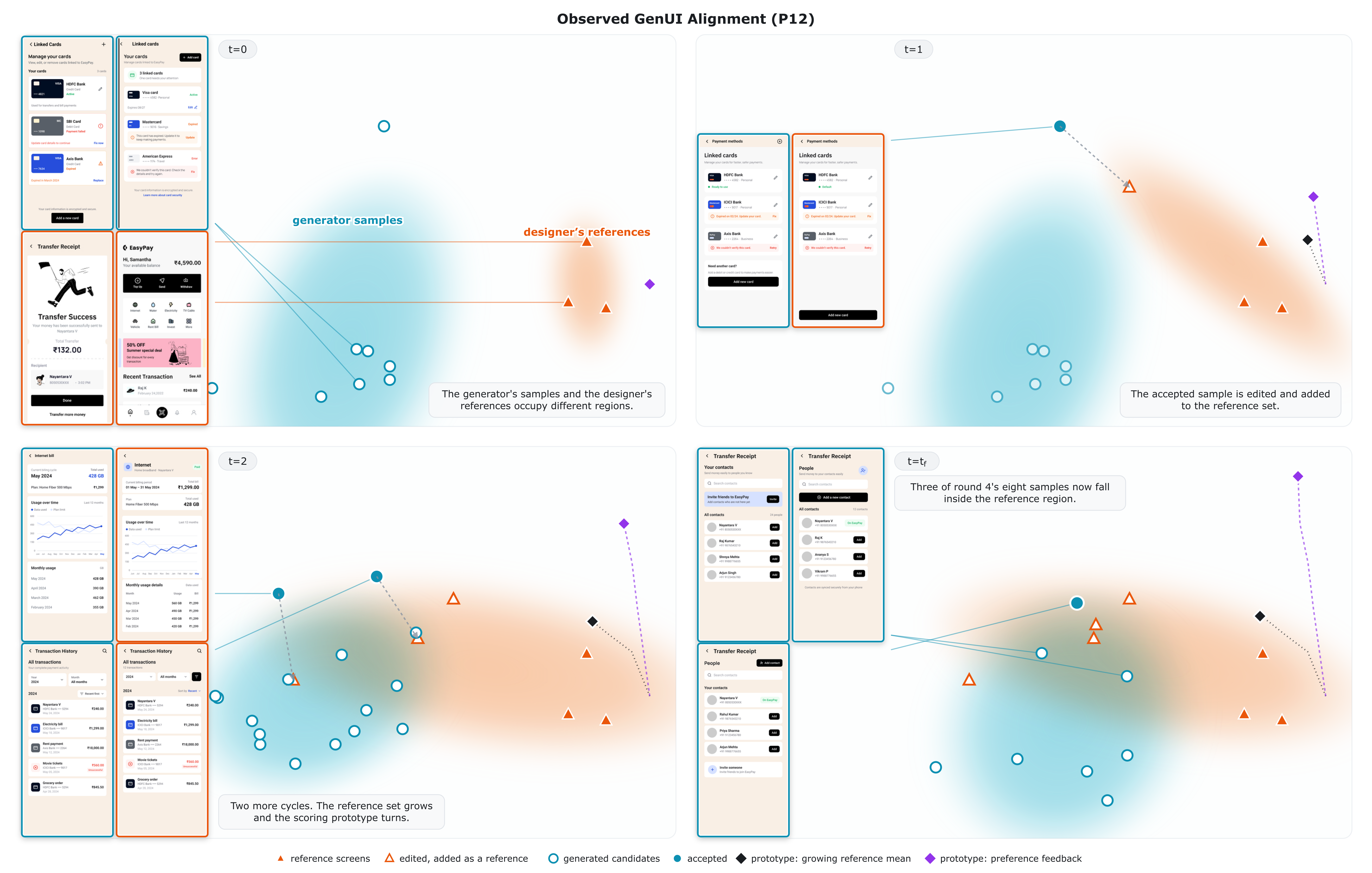}
    \caption{Observed alignment behavior of P12 matches our conceptual model. 
    To generate this visualization, we projected our scoring model's embeddings on actual on-canvas designs retrieved from P12's study logs. Over the course of four rounds, P12's input guided the model's output distribution (blue) towards artboard references (orange) and updated the design system prototype. Note that we cannot observe the true output distribution of the generator and tacit definition of designers, so for the purpose of this visualization, we plotted the covariance ellipse corresponding to $2\sigma$ of mass. Also, note the initial prototype is computed using cosine similarity, which does not result in it being visually centered.}
    \label{fig:conformance-observed}
    \Description{This four-part scatter plot diagram, titled "Observed GenUI Alignment (P12)," illustrates the alignment process over four time steps. At t=0, the generator's samples (a blue region with circles) and the designer's references (an orange region with triangles) occupy entirely separate areas. At t=1, an accepted blue generator sample is edited and added to the orange reference set. By t=2, after two more cycles, the orange reference set has grown, the scoring prototype shifts, and the blue generator distribution moves closer. By the final round (t=t_f), the blue generator samples heavily overlap the orange reference region. The accompanying caption clarifies that the shaded regions represent the covariance ellipse corresponding to $2\sigma$ of the probability mass, used here because the true output distribution and the designer's tacit definitions cannot be directly observed.}
\end{figure*}
\subsubsection{Usage Statistics}
We analyzed interaction logs to characterize how participants used GUIDE during the study.
Across the entire study, participants submitted 95 unique prompts averaging 44 words (one prompt was dropped due to a network error).
In total, participants made 2,619 editing actions and, on average, spent 3.3 minutes making 27.6 editing actions on each screen.
Usage also varied substantially across participants, with some participants spending significantly more time crafting prompts and selecting variants while others spent significantly more in manual editing. For example, P1 spent the most time refining model-generated variants, making an average of 51.1 edit actions, and P8 spent more time on prompting and variant selection, making only 7.4 edit actions on each generated screen.
Our results suggest that GUIDE is able to accommodate these different interaction preferences in design workflows.

\subsubsection{Arena Evaluation Results}
The results of participants' arena evaluations are shown in Figure \ref{fig:arena-results}.
We used the arena evaluation to assess whether participants successfully aligned the GenUI so that they would prefer its outputs over those from the initial baseline GenUI. After each session, participants completed 10 blind comparisons, yielding 240 comparisons overall. One comparison was marked as a tie and was excluded.

Overall, participants significantly preferred designs generated from their own aligned GenUIs, and they selected aligned outputs in 61.5\% of decisive comparisons (p < 0.001). At a participant level, 8 of 12 participants favored the aligned GenUI, 3 were tied, and 1 favored the initial GenUI, which was significant via a sign test (p=0.04).
Participant-led alignment also improved conformance to all artboards, i.e., average win-rate of over 50\%, although some artboards were more difficult.
Specifically, the Digital Payments App artboard led to the lowest arena win-rate average (53.3\%) compared the Learning Management App (65.0\%), Fitness Tracking App (61.0\%), and Food Delivery App (66.7\%).
Overall, these results suggest that iterative designer alignment improves conformance beyond what can be achieved by a GenUI configured once from static design artifacts.

\subsubsection{Scoring Model Improvement}

While the arena evaluation measured alignment at the end of each session, it did not show how this alignment developed over successive iterations of feedback. To this end, we examined each variant selection as a repeated judgment of conformance. We focused on the scoring model because these selections directly evaluate its rankings, since other components of the GenUI has no similar equivalent intermediate signal. As a measure of scoring performance, we report pairwise ranking accuracy, which measures how often the variant selected by the user scores above an unselected alternative, e.g., random chance would be 50\%.

We used two protocols for computing this pairwise accuracy metric over the course of a user's session. The \textit{sequential protocol} scores each iteration once using only the feedback available beforehand, matching how the model is used during deployment. However, because the evaluated screens change across iterations, this also reflects differences in the difficulty of each round's target prompt. Therefore we collected all variants corresponding to rounds 2-4 (with at least one round of alignment) and evaluate each round's scoring model on the entire set. We refer to this as the \textit{fixed-target protocol}.
Using each protocol, we tested the two methods (Section \ref{sec:model-architecture}) for updating the scoring model's prototype vector based on the feedback it received at the time.
The results of our experiments are shown in Figure \ref{fig:overall-scorer}.

Across both protocols, preference feedback outperformed reference averaging as more feedback accumulated. Reference averaging produced only modest gains in the sequential protocol and no improvement in the fixed-target protocol, where it eventually performed worse than using the initial references alone. This differs from our offline AMEX experiments, where additional references generally improved performance, suggesting that the benefits of reference averaging do not necessarily carry over to interactive alignment. 
On the other hand, preference feedback continued to improve with additional interactions, achieving a final pairwise accuracy of 64.7\% under the sequential protocol and 59.6\% under the fixed-target protocol. This suggests that participants' implicit rankings provided more useful information than reference expansion.

The embedding trajectory for P12 provides a concrete view of how this alignment developed during a session (Figure~\ref{fig:conformance-observed}). Initially, the generated candidates occupy a different region from the participant's reference screens, despite being prompted with them, indicating flaws in the system's interpretation of this partial evidence. As the participant selects and refines candidates, the resulting screens are added as references, the scoring prototype shifts, and subsequent candidates increasingly overlap with the reference region. By the final iteration, the two regions substantially overlap, closely resembling the process illustrated in Figure~\ref{fig:conformance-concept}. Together, these results show that repeated feedback from participants could successfully shift a GenUI's output distribution towards their design intent.

\begin{tcolorbox}[
    colback=guidelightblue!35!white,
    colframe=guideblue!20!white,
    boxrule=0.4pt,
    left=6pt,
    right=6pt,
    top=4pt,
    bottom=4pt,
    arc=1pt
]
Our quantitative results showed that GUIDE effectively incorporated designer feedback to align the underlying GenUI with tacit design intent. Final aligned outputs were significantly preferred over the untuned model, while intermediate measurements showed alignment improving as feedback accumulated.
\end{tcolorbox}

\subsection{Qualitative Results}

We analyzed participants' post-study interviews to understand how they i) judged generated variants, ii) experienced GUIDE's adaptation to their feedback, and iii) how the system fit their design workflow.

\subsubsection{Selection Criteria}
Given that most of our quantitative results were derived from participant selections, we sought to understand the underlying criteria participants used to make those selections.
In our interview, we asked participants to name factors that influenced their variant selection choices (i.e., within the plugin workflow).
Seven participants noted that first checked whether the candidate included the information and content they specified in the prompt (P1, P4, P5, P7, P9, P10, and P11).
Given that some participants wrote very detailed prompts (P5 wrote up to 105 words in a prompt), it is possible that variant selection may have been influenced by faithfulness to detailed requirements rather than solely on design system conformance.
On the other hand, P11 also mentioned that they considered factors that were not directly specified in their prompt. For example, P11 recalled that they specified that a teacher profile screen should contain a rating and inspected candidates based on whether the way that information was presented (e.g., filled stars vs bar graph) matched their imagined implementation.
This observation supports our conceptualization that design intent is often only partially specified and requires designer feedback to accurately evaluate.

Nine participants explicitly mentioned that they evaluated candidates based on how closely they matched the reference designs.
Participants recalled checking specific visual properties to more abstract characterizations. For example, P3 noted that they checked whether elements from the references, such as rounded rectangles, were carried over into generated screens.
P4, P6, P7, and P10 considered whether the overall visual style felt consistent with the references.

Finally, many participants also mentioned that they incorporated factors beyond visual conformance in making their selections.
Seven participants discussed usability or user experience, including accessibility (P2, P12), navigation (P8, P12), and whether the screen structure made information easy to understand (P1, P2, P5, P6, P9, P12). Some participants (P2, P7, and P11) noted that one consideration for selecting variants was the amount of work required after selection to refine it. P11, for example, preferred variants that already handled difficult-to-build elements well so that less manual editing would be needed afterward. This suggests that selection did not always track visual conformance alone. Participants could prefer a less conformant variant if its remaining differences were easier to fix, such as a screen with the wrong background color that could be corrected with a single property change.

\subsubsection{Feedback and Adaptation}

Similar to the quantitative results from our scoring model evaluation, participants generally reported that they felt the GenUI became more conformant to their intent over the course of a session.
When asked whether generations became more conformant across successive iterations, participants gave a mean rating of 7.42 out of 10 (SD = 1.08, range = 6--10). Eight participants reported noticing their adjustments were reflected in later generations that needed less manual correction over time.
For example, P5 recalled manually changed a button and observed the same change in the following generated variants.
P4, P5, P7, and P9 all mentioned that screens they generated during the study matched the provided references well, and P3 felt that GUIDE allowed them to more conformant designs than existing AI tools for UI generation.

Participants also noticed cases where their feedback was not incorporated as expected. P4 reported repeatedly making edits that were not reflected in later generations, while P5 wished the system could ``consider what changes I am making more directly.'' Several participants suggested more explicit ways to communicate what should persist. For example, P1, P4, and P7 mentioned that it would be useful to accompany edits with textual comments, and P7 and P9 similarly suggested defining persistent guidelines in advance rather than repeatedly correcting the same issue.
Finally, participants also suggested additional modalities beyond text and direct editing such as sketches (P6, P10, P11) and context-specific feedback (P12). Prior work shows that other forms of designer feedback can also support model training~\cite{wu2026improving}, suggesting broader feedback channels for future designer-in-the-loop systems like GUIDE.

\subsubsection{Design Workflow}

Finally, we asked participants whether GUIDE's alignment interactions complemented or impeded their normal design workflows. Based on our conceptual model of GenUI alignment, we hypothesized that generating multiple variants is primarily useful for designers inspect and estimate the GenUI's output distribution.
Participants (P1, P4, P10) reported that they found the variants useful for exploring or developing new design ideas, suggesting utility in a traditional design workflow.

Other participant suggestions pointed towards more flexibility in applying refinements.
P2, P3, and P5 felt that it would be more efficient to make refinements through additional rounds of prompting rather than requiring direct manipulation.
One potential factor identified by P4 and P6 was that direct manipulation editing was harder on many variants since they were converted from LLM-generated HTML/SVG and lacked familiar Figma structures such as auto-layout and organized containers. Overall, these findings suggest opportunities to refine both GUIDE's interaction design and our underlying conceptualization of GenUI alignment.

\begin{tcolorbox}[
    colback=guidelightblue!35!white,
    colframe=guideblue!20!white,
    boxrule=0.4pt,
    left=6pt,
    right=6pt,
    top=4pt,
    bottom=4pt,
    arc=1pt
]
Our qualitative results showed that participants made selections during both design and evaluation based on factors beyond conformance to the reference artboards. Participants nevertheless felt that GUIDE incorporated their design intent effectively and complemented their familiar design workflows.
\end{tcolorbox}
\section{Discussion}
Our results show that GUIDE can help designers align GenUI behavior with their design intent. In this section, we discuss the implications of this work for generative design. We first consider the new gulfs designers face when evaluating and shaping generative behavior, then examine how conformance is negotiated in practice. Finally, we discuss how familiar methodologies in HCI can be applied when designing dynamic, ephemeral artifacts.

\subsection{Gulfs of Generative Design}

In their paper, Hutchins, Hollan, and Norman describe two ``gulfs'' at the root of many usability challenges~\cite{hutchins1985direct}. They described i) the ~\textit{gulf of evaluation} between system state and what users can understand and ii) the \textit{gulf of execution} between user goals and the actions available to achieve them~\cite{hutchins1985direct}. These gulfs were originally framed around end users, who could not directly inspect the systems they used, rather than around designers who faced less of this uncertainty because they could usually inspect the artifact they were building. However, GenUIs and generative design make both gulfs relevant to designers as well, since designers often lack complete visibility and control over these systems.

Hutchins et al.~\cite{hutchins1985direct} proposed direct manipulation as a way to reduce these gulfs by letting people act directly on representations of data or objects. GUIDE brings similar ideas to GenUI design by providing an environment for direct manipulation-driven editing. First, GUIDE seeks to make the output distribution of the GenUI more readily visible through repeated and readily available generation, reducing the gulf of evaluation of system behavior. On the other hand, GUIDE reduces the gulf of execution by letting designers make the desired change directly on the generated on-canvas outputs, which are represenatative of the GenUI's internal state and are artifacts used for future generations.

Our study provides evidence that GUIDE successfully operationalizes the gulfs of evaluation and execution for generative design in the form of a usable tool. Overall, participants were not only able to produce GenUIs that better aligned with their design intent, most felt that GUIDE helped me achieve this goal complemented their existing design workflows. To summarize, generative systems make uncertainty about system behavior a problem for designers, not just end users, and GUIDE represents one way to apply longstanding HCI principles for bridging these gulfs to generative design.

\subsection{Negotiating Conformance}
The conceptualization of conformance described in this paper guided the design of both the scoring model and GUIDE, and in our study, we observed some evidence to validate aspects of this model (Figure~\ref{fig:conformance-observed}). At the same time, other observations suggest that this simplified picture leaves out important nuances of how conformance is negotiated in practice.

First, our qualitative results already highlighted that participants' selection criteria are based on more than conformance alone, and ranged from adherence to their prompt (e.g., P5), overall UX factors (e.g., P2, P12), and amount of effort needed to make corrections (e.g., P11). This observation suggests that in practical settings, conformance to some imagined design intent must also be negotiated against different factors. Some work~\cite{wu2024uiclip} has already investigated building models of these principles, and future tools for GenUI alignment might incorporate multi-objective optimization approaches or Pareto-aware selection in generating variants~\cite{dayama2020grids}.
\begin{figure}[!htb]
    \centering
    \includegraphics[width=17pc]{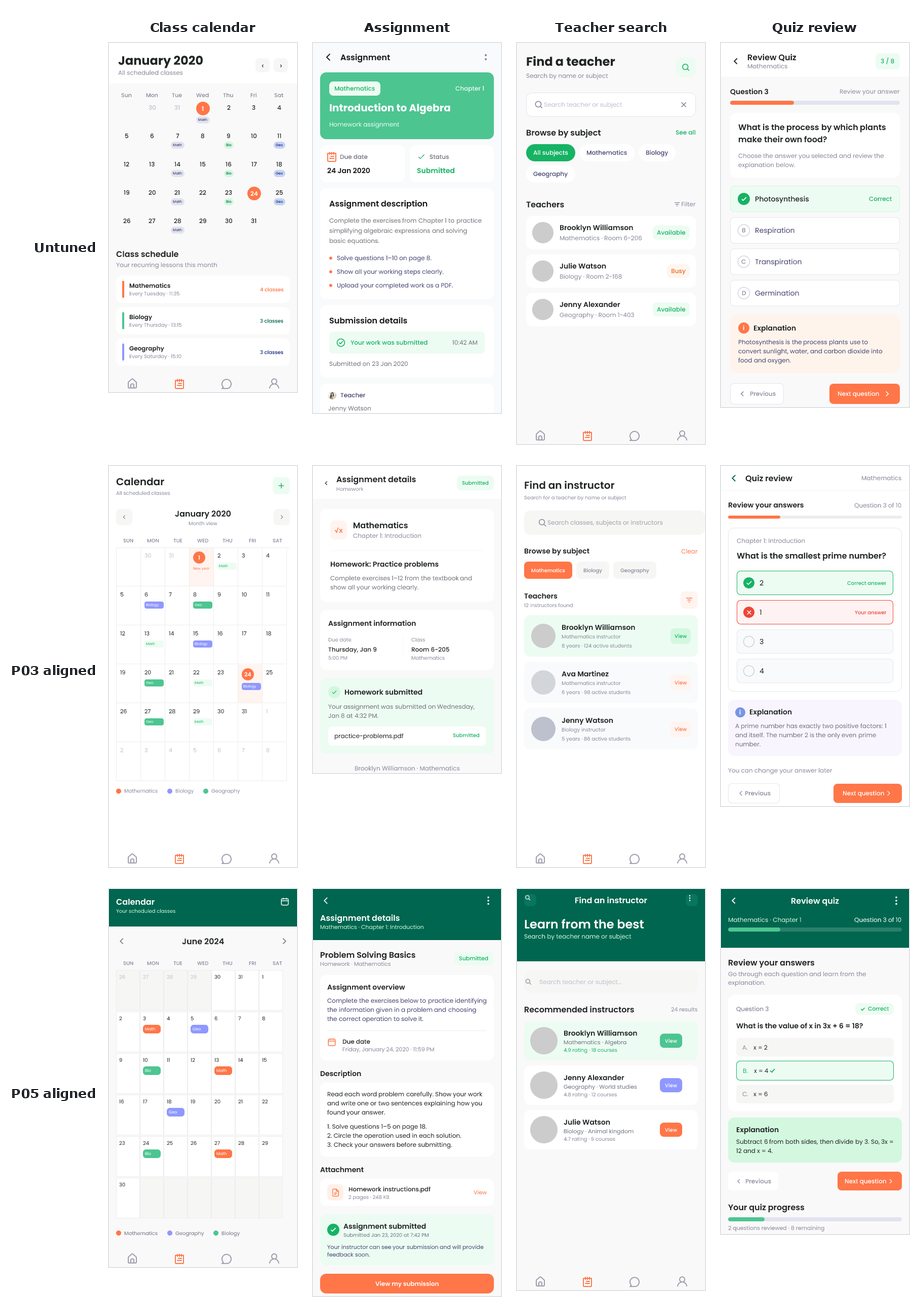}
    \caption{P3 and P5 started from the same materials and untuned GenUI for the Learning Management App but aligned them substantially differently. Figure shows screens generated using untuned (top), P3's tuned (middle), and P5's tuned GenUI on held-out arena test descriptions. Unlike the Untuned GenUI, both aligned GenUIs are internally consistent. Aligned GenUIs showed different tendencies from each other, where P3's adopted a lighter theme and multi-column action buttons and P5's prioritized a dark green header and full screen elements.}
    \Description{Three rows of screenshots generated by three different GenUIs, an untuned generator, one aligned by P3, and one aligned by P5. Unlike the Untuned GenUI, both aligned GenUIs are internally consistent. Aligned GenUIs showed different tendencies from each other, where P3's adopted a lighter theme and multi-column action buttons and P5's prioritized a dark green header and full screen elements.}
    \label{fig:useralignments}
\end{figure}
In addition, we found that different participants could interpret the same references differently. In our study, participants were not the original designers of the artboards and they themselves had to infer the intended design language (of the original artboard author) from a small number of examples. We observed that this led to varying notions of conformance. Figure~\ref{fig:useralignments} shows GenUIs aligned from the same reference artboards but using feedback from different participants. This observation raises implications for how multiple designers might work together to design GenUIs in a collaborative setting.

Finally, the designer's own notion of conformance could change during alignment. Several participants (P9) reported that generated variants helped them develop ideas they had not considered initially, and several others (P1, P4, P10) also mentioned the benefit of multiple variants not just for gauging generation performance, but also for creative exploration. This suggests that a designer's notion of conformance might be influenced by model outputs. Other suggestions from participants also point to implications for our conceptualization of GenUI alignment. For example, many participants (P1, P3, P9, P10, P11, and P12) felt that selecting only one variant to refine was disruptive, as they often felt different candidates had different strengths and wanted to combine parts of them together. This suggests that the most useful target for updating the GenUI may be a design assembled from multiple outputs rather than any single sample it generated.

Overall, these observations suggest that a designer-in-the-loop formulation is appropriate for editing and refining these evolving definitions.

\subsection{The Design of Ephemeral Things}

Many HCI design methodologies, such as sketching~\cite{buxton2007sketching} and the Double Diamond~\cite{designcouncilDoubleDiamond}, describe processes that begin with exploration and ``narrow in'' toward a solution over time. Our work investigated this problem specifically in the context of GenUIs, which are increasingly appearing in end-user products~\cite{leviathan2025generativeui,anthropic2025artifacts}, but raise challenges for these existing methodologies due to output unpredictability. While our work primarily develops models and tools for GenUIs, similar challenges arise when applying existing design methodologies to other generative applications.

Related work in HCI has studied design in similarly dynamic contexts. For example, service design develops methods for shaping experiences that may unfold differently each time~\cite{bitner2008service}. These methods rely on relatively persistent representations such as personas, journey maps, and service blueprints~\cite{stickdorn2018service}. Similar practices are emerging in AI design, where designers refine prompts and evaluation cases to steer and measure model behavior across many possible outputs~\cite{mather2026modeldesigner}.

In our work, we built a tool that combined similar persistent representations for GenUIs, including prompts, exemplars, and learned representations. While the final design output of the GenUI could vary, designer used our tools to observe and refine relatively stable targets that made behavior more predictable. Our study results showed that most participants (8/12) felt that their edits were reflected in subsequent generations, and our scoring model evaluation provide additional analytical evidence that feedback accumulated through these representations from measured performance improvements to inspections of model internals. 

To summarize, designing GenUIs, AI systems, and other ``ephemeral'' artifacts presents unique challenges for HCI practitioners and methodologies. Our paper presents a tool that gives designers persistent representations through which they can inspect and shape behavior even when individual outputs are transient and uncontrollable. We hypothesize future conceptual formulations and tools will be useful in bridging familiar design processes new domains.

\section{Limitations \& Future Work}
While we show that GUIDE is an effective tool for aligning GenUIs, we acknowledge that our system and study has several limitations. In this section, we detail opportunities for improving these limitations and address opportunities for future work. 
\paragraph{Technical Limitations.}
While some previous work has approached generative alignment in the context of finetuning a model's weights~\cite{wu2024uicoder,wu2026improving,ouyang2022training}, we adopt a heterogenous definition of a generator system that contains a fixed base generator, prompts, and a separate scoring model for inference-time guidance. We made this decision to support rapid updates in our designer-in-the-loop setting, since updating LLM weights would be computationally expensive and time-consuming. In a production version of our system, we envision a feature that could ``bake'' alignments into a self-contained model by distilling this heterogeneous system's data into a exportable model. Another limitation of our work was that the conformance scorer operates only on rendered screenshots, while style data such as exact color hex codes and margins are available in the Figma environments. This decision was driven in part to the lack of style metadata in our AMEX pretraining dataset. During development, we investigated building a model from a smaller SVG dataset but did not reach the scale of data needed for robustness. We leave this modeling opportunity to future work.

\paragraph{User Study Limitations.}
Our user study was limited by several practical constraints. First, participants designed only four screens per app on a limited set of artboards, which might not be fully representative of real-world workflows. A future study could examine a broader range of artboards or allow participants to bring their own, although we used a fixed set here to control variation across participants. Due to the 35-minute limit of our sessions, we were unable measure longer term phenomena, as is typical for real UX development~\cite{getto2022triuxpa}. Finally, we focused on evaluating GUIDE as an integrated system rather than ablating its components, e.g., effect of prompt optimization vs scoring model improvement, since our primary goal was to evaluate how commonly used GenAI alignment mechanisms work together in a realistic workflow.
Although we separately analyzed some components (e.g., scoring model), a full ablation would have required generating and collecting designer judgments for significantly more comparisons, which would exceed our study's time constraints.

\paragraph{Future Work from Discussion Implications.}
Finally, our discussion points to several promising directions for future work. The gulfs of generative design point to additional opportunities for translating learnings from end-user systems to design tools for generative AI. Our findings on negotiating conformance suggest that alignment systems should treat conformance as an evolving target, potentially leading to an updated conceptual model and computational formulation that models the designers' own shifting conformance definitions. Finally, our discussion of ephemeral interfaces suggests that future work should examine what persistent representations designers can refine when the generated interfaces themselves may exist only temporarily.

\section{Conclusion}
In this work, we introduce GUIDE, a designer-in-the-loop system that aligns GenUI systems toward generating more conformant screens through iterative cycles of inspection, generation, and modification. We formalize design conformance as shifting a GenUI's output distribution toward a designer's tacit conformant set, and operationalize this with a scoring model that achieves up to 96.7\% macro accuracy in representing design identity and matches or outperforms a proprietary LLM baseline in predicting conformance relationships. In a study with 12 UI/UX practitioners, participants significantly preferred designs from their GUIDE-aligned GenUI over a strong baseline. These findings suggest that designer-in-the-loop feedback offers a rich, reliable signal for alignment that static examples or guidelines alone cannot fully capture, and point toward a future where GenUIs can be deployed to generate outputs consistent with designer intent.

\bibliographystyle{ACM-Reference-Format}
\bibliography{reference}

\clearpage
\onecolumn
\appendix

\section{Implementation Details of GUIDE}
\label{sec:technical-details}


\subsection{Prompt for \texttt{design.md} Synthesis}
\label{sec:prompt-1}

\begin{promptbox}
You maintain a design system reference document. Below are facts about one design system, measured directly from a real Figma file named "{fileName}" -- exact colors, exact typography, exact spacing, exact components. Rewrite this into a well-organized document a designer would actually want to read, using only the facts given.

--- EXTRACTED FACTS ---
{factsDocument}
--- END EXTRACTED FACTS ---

{if hasImages: "Screenshots of the reference screens are attached, for judging visual character only -- overall mood, density, how photography or illustration is used, how loud or quiet the typography reads in practice. Do not state a color, size, or any other value because you can see it in a screenshot; state it only when it also appears in the facts above. A screenshot tells you how something is used, never what its exact value is."}

What you may change from a flat list of facts:
- Group related values under your own subheadings (a "###" heading inside a "##" section) and explain what each is for, when the evidence actually supports a specific role -- which color is a primary action versus a hairline border, which text style is for a heading versus a caption. Base every explanation on the facts or the screenshots, never on a guess about what a typical app of this kind usually looks like.
- Under each heading, add a short descriptive line for a value when the Component Specs section (or another part of the facts) actually ties it to a specific place it is used -- e.g. "used as the fill on Button/Primary and the active tab indicator" for an accent color, or "used for section titles; Body/Regular is the more common text style, appearing on most cards" for a typography style. Point to the specific component, screen role, or other fact that shows this, not a general impression from the screenshots. If the facts don't tie a value to any particular use, leave it as a plain token instead of inventing a usage.
- You may write one "## Do's and Don'ts" section, listing the clearest of those same usage facts together in one place -- "- Do: use primary-purple as the fill for primary actions (Button/Primary, active tab indicator)." Same standard as the line above: only a usage a specific fact actually ties a token to, never a general impression. This section is Do only. Never write a "- Don't:" line here -- a Don't can only ever come from a designer's own correction, moving away from something the document said, and there is no correction history at this stage for one to be grounded in. Skip the section entirely if nothing in the facts ties a token to a specific use.
- You may write one "## Overview" section, placed first, of 2-4 sentences on the overall character: how restrained or bold the palette is, how the type scale gets used, anything visually distinctive about these specific reference screens. Write only what these specific facts and screenshots actually support -- this is not the place for generic design commentary.
- Where the facts genuinely leave something undetermined, say so in a line or two rather than guessing at a plausible-sounding answer.

What must not change:
- Every "##" heading you use, other than "Overview", must be copied exactly from this list, and only for ones that have content: {DESIGN_MD_SECTIONS + DESIGN_MD_PROSE_SECTIONS, e.g. "Colors, Typography, Spacing, Radius, Component Specs, Do's and Don'ts"}. Keep them in that order.
- Never state a color, font size, spacing value, radius, or any other number that is not copied exactly from the facts above. This document is checked afterward against those exact facts, and the whole thing is discarded if anything in it is not traceable to them.
- Never invent a component, a variant, or a behavior -- a hover state, a breakpoint, an animation, anything responsive -- that the facts do not describe. If the facts say nothing about how this adapts to a different screen size, do not write a section about it; there is nothing to base one on.
- Never mention Figma, a layer name, a node id, or any other detail about how these facts were extracted. Whoever reads this document will never see the Figma file it came from.
- Never name or compare this to any other real brand, product, or company. Describe only what is actually in front of you.
- Return only the rewritten document as Markdown. Do not explain what you changed or add commentary outside the document itself.
\end{promptbox}

\subsection{Prompt for desgin.md update}
\label{sec:prompt-2}

\begin{promptbox}
You maintain a design system document. A screen was generated from it, the designer corrected the parts they did not want, and then accepted it. Each correction is a place the document told the generator something the designer disagreed with.

Here is the document:

--- CURRENT DESIGN SYSTEM ---
{designMd}
--- END CURRENT DESIGN SYSTEM ---

{correctionBlock: WHAT THE DESIGNER CHANGED -- property diffs, added/removed elements, comments}
{structureBlock: before/after screen-structure trees, node name + pixel rect}
{if two images attached: "Two images are attached: the first is the screen as the document produced it, the second is the screen after the designer corrected it. The corrections above are not the complete record of what changed -- they can only ever describe a limited set of properties (fill, size, radius, text, font size, among a few others) on an element that already existed when the round started. Two whole categories of change never appear there at all, no matter how deliberate: a property outside that tracked set (...), and an element being added, removed, or swapped for a different one (...). These two images are the only complete record of the screen; compare them directly for anything like this..."}

Do not return the document, and do not describe the changes in your reply. Make them by calling the edit_design_md tool -- once per change, old_str the existing line you found it at (copied exactly, character for character, from the document above), new_str what it becomes.

To change an existing value: old_str is the line exactly as it reads now, new_str is what it becomes.
To add a new fact without touching anything existing: old_str is any nearby existing line (a heading anchors the addition as that section's first entry), and new_str is old_str's own text followed by a new line for the addition -- so what you send includes the anchor unchanged, plus what's new.

[... YAML-block rules, disambiguation-by-heading rule, and four worked examples: (1) an existing value changes deliberately, (2) a new fact anchored to an existing line, (3) a "Don't" recorded only from an actual correction moving away from something specific, (4) naming a new fact by the element's ROLE (screen title, primary action) rather than by this screen's own content label, (5) resolving an ambiguous old_str with a two-line section-heading anchor ...]

Rules:
- Before deciding anything, ask why the designer likely made each correction -- not just what value changed. A value is only evidence for the document when it looks like a deliberate design choice the designer would want to hold on other screens too. A value that only fits this screen's specific content (a box resized because its text got longer, a number changed because the copy changed) is not that, however precisely you know the before and after -- do not record it. When you cannot tell the difference, say so by leaving that correction out rather than recording it anyway.
- Then ask what general pattern the correction is evidence for, not only the number it happened to produce: a single measurement is not itself a design rule, a relationship between that measurement and what it's for might be, if that is what a deliberate correction actually supports. Prefer writing the pattern you can actually infer over the raw measurement. When no general pattern is inferable -- the evidence is one number on one screen and nothing more -- record only that, plainly; do not invent a rule the evidence does not support.
- Preference between two general treatments (not just one number replacing another): record it as a "Don't" under "## Do's and Don'ts", naming what was moved away from, not only the destination. Never write a "Don't" that is not grounded in an actual correction moving away from something specific.
- "old_str" must be a line that appears in the document above, copied exactly, character for character.{+ note on missing sections, if any}
- old_str ambiguous across sections: retry once with old_str as two lines (the section heading, then the bullet), never fall back to an unanchored new bullet instead.
- A correction that moved a value away from one the document specifies, and reads as a deliberate choice rather than a fit to this screen's content, means the document is wrong about that value: change it to what the designer chose.
- A correction to something the document does not mention, and that reads as a deliberate choice, means the document is silent where it should not be: add what the designer chose, anchored to the line it belongs closest to.
- Change the most specific bullet the correction is about, and only that one. A correction to one component belongs in that component's spec. The same value may also appear as a shared token, and a single component changing is not a reason to redefine the token for everything that uses it -- change the token only if the correction is plainly about the token itself.
- Naming a brand-new fact: name it by the element's role, never by this screen's own content.
- When you change a bullet, alter only the part that changed. The rest of the bullet must come back exactly as it went in.
- Change only what a correction is evidence for. The rest of the screen was not corrected and says nothing about the document.
- A correction that is clearly about this screen's content -- wording, a number, which item is selected -- is not a design system change.
- new_str must never be empty. There is no way to remove anything and you should not want one.
- Never introduce a value that appears in neither the document nor the corrections.
- One correction is one designer changing one thing once. Calling the tool zero times is a reasonable outcome.
\end{promptbox}

\subsection{Prompt for screen generation}
\label{sec:prompt-3}

\begin{promptbox}
{if reference images attached: "These are reference images from an existing design system. Match their visual style -- colors, typography, spacing, iconography, and component styling -- but follow the layout, content, and structure described in the request below. Only mirror the references' layout or composition where the request does not specify its own.\n\n"}Generate a UI component as an HTML fragment (target size {width}x{height}) for: "{prompt}". {bgNote: "Always include a full-size background rectangle as the first element with a color matching the reference image's background." | "Always give the root element a full-size white background."}

{designContext: if design.md exists -- "Follow this design system strictly for colors, typography, spacing, and component styles:\n{designMd}" (normal case), OR when re-exploring the same screen: "Use this as your visual foundation for colors, typography, and spacing -- see the note below about introducing your own variation on specific components:\n{designMd}"}

{priorityNote, only when both reference images AND design.md are present: "If there is any conflict between the reference images and the design system, prioritize the design system -- for structural and compositional choices (which control fills a given role, such as a back icon versus a profile photo in a header's leading slot) exactly as much as for colors, sizes, and spacing. A reference image showing the old pattern does not outweigh a design system that has since been corrected away from it."}

{newScreenNote, only for round 2+ describing a NEW screen: "The reference images show other screens from this same product, provided so you match their design system -- colors, typography, spacing, iconography, and component styling. Do not copy their layout or content: build the screen described in the request above, using their visual language."}

{componentInstruction, only when the design system has catalogued reusable components: "DESIGN SYSTEM COMPONENTS -- these repeat across the reference screens, so they belong to the product rather than to any one screen: {specs}. This is what is available, not a list of things to include. Decide from the request what this screen actually has ... For the parts the screen does have, use these rather than inventing equivalents. There are two correct ways: 1. UNCHANGED -- emit only this placeholder, with nothing inside it: <div data-use=\"NAME\" style=\"display:flex;width:WIDTHpx;height:HEIGHTpx\"></div> 2. DIFFERENT CONTENT -- copy its markup below and edit only the content ... Navigation bars, tab bars, status bars and home indicators carry no screen-specific content, so when the screen has one, use option 1 for it. ... A part marked \"reserve-only\" has no text in it ... Do not draw your own version of anything on this list. {componentMarkup}"}

This HTML will be rendered by satori, not a browser. Follow these rules exactly:
- Return only one visual root <div>; do not include <!DOCTYPE>, <html>, <head>, <meta>, <title>, or <body>.{if components listed: "\n- Never draw your own navigation bar, tab bar, status bar or home indicator when one is listed as a design system component above -- emit its placeholder div instead."}
- Every style must be inline. Use flexbox only; do not use CSS Grid, floats, absolute positioning, classes, style blocks, or external stylesheets.
- justify-content accepts only: center, flex-start, flex-end, space-between, space-around. Never use space-evenly -- satori rejects it and the whole variant fails to render.
- Every <div> and <span>, including empty ones, must have an explicit display property. Use display:flex in almost every case and set flex-direction:column when children should stack.
- Every element that directly contains text must have an explicit numeric line-height (e.g. line-height:1.3 or line-height:20px), not just font-size. Do not omit line-height on any text element, including titles and headings. satori miscalculates the box height of text with no explicit line-height, which causes it to overlap the next sibling element instead of stacking cleanly below it.
- Use only numeric CSS lengths such as 12px, 1.25, or 100%; never spell out a number in a numeric property.
- Any text that should be centred must carry text-align:center on the text element itself, not only justify-content on its container. Centring a full-width heading or button label through the container alone leaves the text pinned left once imported.
- Everything must fit inside {height}px of height. Count it as you go: a screen that runs past the bottom does not scroll here, it collides with whatever sits at the foot of the design -- the tab bar, the home indicator. If the content will not fit, cut a section or shorten the copy rather than letting it run over.
- Never put a fixed width or height on an element that contains text, or on the pill, badge, chip or button wrapping it. Size those with padding and let them take the width of their contents. A fixed box smaller than its label does not shrink the text -- the text spills out of it and collides with whatever is next to it. Fixed sizes are for spacers, images, icons and the screen root only.
- Set font-family explicitly on text elements, using the exact font families named in the design system above (list a generic fallback after it, e.g. font-family: 'Inter', sans-serif). Do not substitute a different typeface. {or, when no design.md: "Leave font-family unset or use sans-serif -- no design system fonts were provided."}
- For photos or avatars, use <img data-image-query="descriptive search query" alt="descriptive search query" style="display:flex; width:120px; height:120px; object-fit:cover;" />. Do not put a remote URL in src; the plugin resolves the query through Pexels before rendering and uses a gray placeholder if unavailable. Add border-radius when the design system's images are not square -- border-radius:50% for a round avatar -- and match whatever radius the design system uses; an image is otherwise imported with square corners.
{textStyleInstruction, only when the design system has a text-style vocabulary: "- Text styles: the design system defines exactly these, and no others: {list}. On every text element that corresponds to one, add data-text-style=\"NAME\" using a name from that list verbatim ... Never invent a name that is not on the list."}
- A standard icon (search, close, chevron, heart, star, checkmark, and the like) that is not offered as a design system component above must be drawn as real inline SVG: <svg width="24" height="24" viewBox="0 0 24 24" fill="none" stroke="#111111" stroke-width="2"><!-- path/circle/line/polyline elements describing the icon's actual shape --></svg>, with explicit width and height attributes matching the size it needs in context, and an explicit hex color on fill or stroke (whichever the icon actually uses) -- matching the color it should read as in context, often the same as nearby text. Never write currentColor, a CSS variable, or any other color keyword that depends on inheritance: this SVG is imported standalone, with no surrounding page for a keyword like that to inherit a color from, so it resolves to nothing and the icon comes in with no visible stroke or fill at all. Never represent an icon any other way either -- not a Unicode symbol or emoji character, not a letter standing in for a shape, and not the name of an icon font or set (Material Symbols, Font Awesome, and the like) written as text expecting a ligature, since none of those render as an icon here; they render as that literal text.
- Use plain solid-color divs for decorative elements and shapes that are not standard icons. Only use a gradient where the design system clearly uses one for that same purpose -- do not add gradients to cards, buttons, headers or backgrounds on your own initiative.

Return only the HTML fragment, with no explanation or markdown fences.{explorationNote, only when re-exploring the same screen across a batch: "This reference reflects a direction the designer has already been refining, not an external design system to copy exactly. Keep the overall visual language (palette, typography, spacing) consistent with it, but do not reproduce it verbatim -- bring a genuinely different, tasteful take on at least one part of the design so this is a real alternative rather than a near-duplicate. For this variant specifically: {one of a fixed set of exploration hints, round-robin per variant index}"}
\end{promptbox}

\subsection{Prompt for model validation}
\label{sec:prompt-4}

\begin{promptbox}
    Review the provided reference designs to understand their underlying design system "
    "and language. Compare Candidate A and Candidate B. Evaluate each candidate's "
    "conformance to the reference set. Output strictly a JSON object with the key "
    "'selected' set to either 'A' or 'B' (to select which candidate is more compliant) "
    "and an optional 'reasoning' field.
\end{promptbox}

\subsection{Scoring Model Hyperparameters}
We provide a list of hyperparameters used for training the scoring conformance model on our design conformance dataset in Table \ref{tab:hyperparameters}.
\begin{table}[H]
\centering
\caption{Training Hyperparameters for Scoring Conformance Model}
\label{tab:hyperparameters}
\begin{tabular}{ll}
\toprule
\textbf{Hyperparameter} & \textbf{Value} \\
\midrule
Batch size & 8 rows \\
Training steps & 4,000 \\
Learning rate & $3 \times 10^{-5}$ \\
Weight decay & $1 \times 10^{-4}$ \\
Hinge margin & 0.5 \\
Reference-set sizes & $\{1,2,3,5,8,16,32\}$ \\
\bottomrule
\end{tabular}
\end{table}


\clearpage
\onecolumn
\section{Study Materials}
\label{sec:study-materials}

The user study consisted of two sets (Set 1 and Set 2), each comprising two sessions. Each session was conducted with a different artboard reference and used its own distinct set of screen keywords and evaluation-arena prompts.
 
 
\begin{figure}[H]
    \centering
    \includegraphics[width=\textwidth]{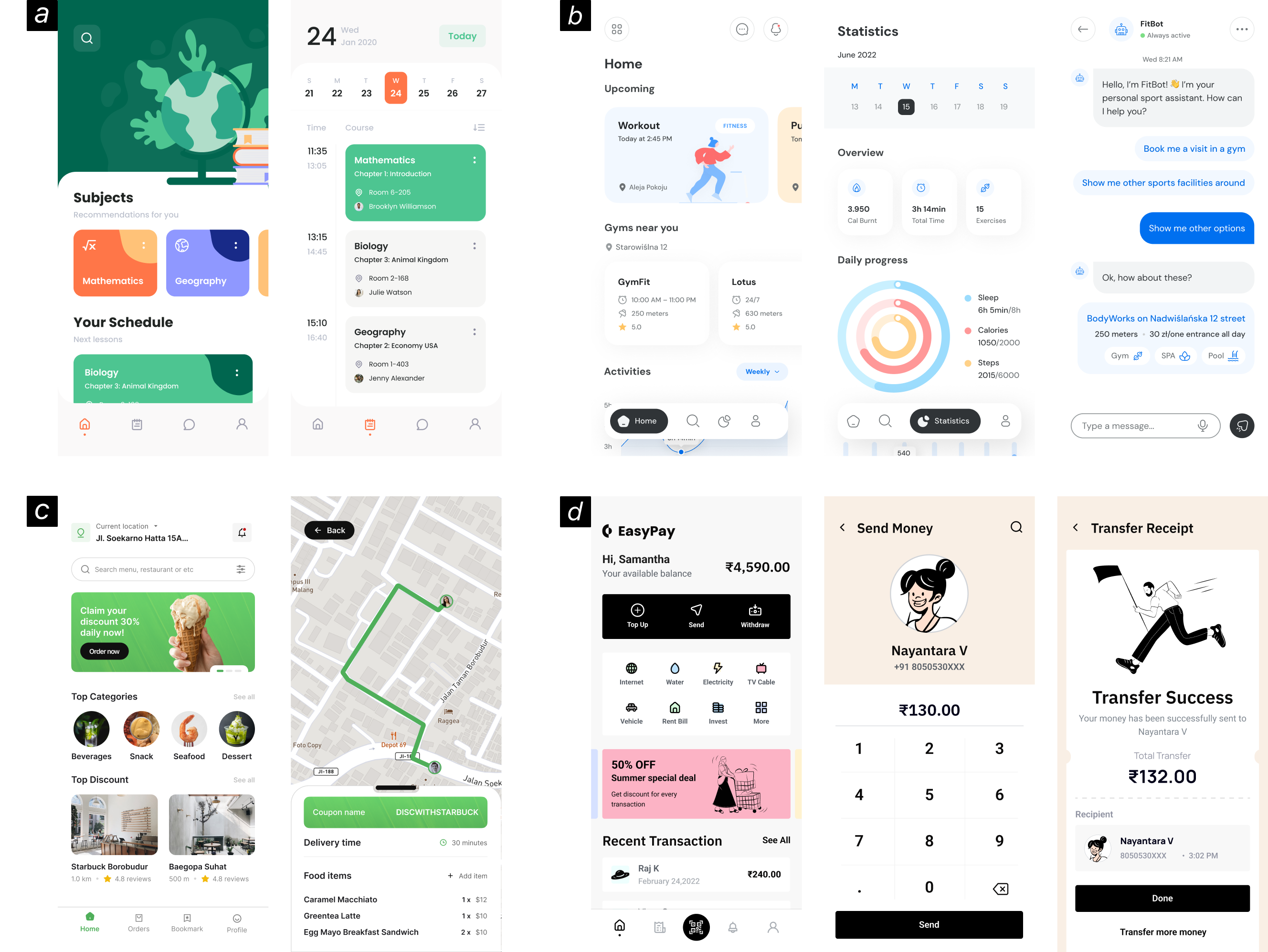}
    \caption{
        Artboard references used in each session of the user study:
        (a) Set 1, Session 1 (Learning Management App);
        (b) Set 1, Session 2 (Fitness Tracking App);
        (c) Set 2, Session 1 (Food Delivery App);
        (d) Set 2, Session 2 (Digital Payments App).
        Each artboard defines the visual and structural reference that
        participants' generated and modified screens were expected to
        conform to within that session.
    }
    \Description{This image displays four sets of mobile UI screens, labeled (a) through (d), which served as artboard references across different sessions of a user study. Panel (a) shows a green-themed "Learning Management App" with scheduling and subject interfaces. Panel (b) features a "Fitness Tracking App" with activity statistics and a chatbot. Panel (c) displays a "Food Delivery App" containing a map, food categories, and order details. Panel (d) shows a "Digital Payments App" highlighting balance, a number pad, and a transfer receipt. The accompanying text explains that these artboards provided the visual and structural benchmarks that participants' generated screens were expected to conform to during the study. Image attribution and licensing details are included at the bottom.}
    \label{fig:artboards-all}
\end{figure}

\begin{center}
\small
\begin{minipage}{0.9\textwidth}
\textbf{Image attribution.} Artboards licensed under
CC BY 4.0 (\url{https://creativecommons.org/licenses/by/4.0/}):
\begin{enumerate}
    \item[(a)] \textit{Schedule Management Platform} (2019) by Ilia Utkin,
    source: \url{https://www.figma.com/community/file/904740192549800638}.
    \item[(b)] \textit{Gym Chatbot App Concept} by Miquido and Filip Rygucki,
    source: \url{https://www.figma.com/community/file/1124291019178940578}.
    \item[(c)] \textit{FreshGo -- Food Apps, Food Delivery App UI Kit} by
    Iko Setiawan,
    source: \url{https://www.figma.com/community/file/1249743721562101377}.
    \item[(d)] \textit{EasyPay: E-Wallet Digital Payment App} by
    Nickelfox Design,
    source: \url{https://www.figma.com/community/file/1146678238901785717}.
\end{enumerate}
\end{minipage}
\end{center}


\subsection{Set 1}
\label{sec:set1}


\subsubsection{Session 1: Learning Management App}
\label{sec:set1-session1}

\paragraph{Screen Keywords Given to Participants.}
\begin{itemize}
    \item Instructor Profile
    \item Class Chat
    \item Search class
    \item Lesson Detail
\end{itemize}

\paragraph{Prompts Used in the Evaluation Arena.}
\begin{table}[H]
    \centering
    \caption{
        Screen keywords and prompts used in the evaluation arena
        for Set 1, Session 1 (Learning Management App).
    }
    \label{tab:set1-session1-arena}
    \begin{tabular}{c p{0.25\textwidth} p{0.6\textwidth}}
        \toprule
        \textbf{Index} &
        \textbf{Screen Keyword} &
        \textbf{Prompt} \\
        \midrule
        1 & Instructor Profile &
        A screen showing a teacher's photo, name, subject, and
        contact/message shortcut. \\
        
        2 & Class Chat &
        A screen showing the 1:1 or group message thread for a
        specific class. \\
        
        3 & Monthly view schedule &
        A screen showing all scheduled classes laid out across a
        full calendar month. \\
        
        4 & Search Instructor &
        A screen where users search for a specific teacher by
        name or subject. \\
        
        5 & Homework Detail &
        A screen showing the full assignment description, due date,
        and submission status for a piece of homework. \\
        
        6 & Review Quiz &
        A screen where students go through quiz questions again
        to review their answers and explanations. \\
        
        7 & Progress Tracker &
        A screen showing a student's completion progress and
        performance across subjects. \\
        
        8 & Class register &
        A screen where students enroll in or sign up for a
        specific class or course. \\
        
        9 & Profile &
        A screen where users view and manage their account info
        and settings. \\
        
        10 & Class review &
        A screen where students rate and leave feedback on a
        completed class. \\
        
        \bottomrule
    \end{tabular}
    \Description{Table 1 lists ten screen keywords and their corresponding descriptive prompts used for evaluating a Learning Management App in Set 1, Session 1. Structured with columns for Index, Screen Keyword, and Prompt, the table outlines the specific UI content requirements for different app features. Examples include an "Instructor Profile" prompt for displaying teacher details and a "Homework Detail" prompt for showing assignment descriptions and due dates.}
\end{table}


\subsubsection{Session 2: Fitness Tracking App}
\label{sec:set1-session2}

\paragraph{Screen Keywords Given to Participants.}
\begin{itemize}
    \item Achievements Dashboard
    \item Exercise Instruction
    \item Payment History
    \item Gym Reservation
\end{itemize}

\paragraph{Prompts Used in the Evaluation Arena.}
\begin{table}[H]
    \centering
    \caption{
        Screen keywords and prompts used in the evaluation arena
        for Set 1, Session 2 (Fitness Tracking App).
    }
    \label{tab:set1-session2-arena}
    \begin{tabular}{c p{0.25\textwidth} p{0.6\textwidth}}
        \toprule
        \textbf{Index} &
        \textbf{Screen Keyword} &
        \textbf{Prompt} \\
        \midrule
        1 & Achievements Dashboard &
        A screen showing badges and milestones earned from
        consistent workout activity. \\
        
        2 & Exercise Instruction &
        A screen showing how to perform a single exercise, with
        animation/video and target muscle group. \\
        
        3 & Reservation History &
        A screen listing past and upcoming gym or facility bookings. \\
        
        4 & Reservation Payment &
        A screen where users complete payment to confirm a gym or
        facility booking. \\
        
        5 & Gym Search Filter &
        A screen where users narrow gym search results by distance,
        price, rating, or amenities. \\
        
        6 & Statistics Leaderboards &
        A screen ranking users by activity metrics like calories
        burnt or workout streaks among friends or the community. \\
        
        7 & Sign Up &
        A screen where new users create an account to start using
        the app. \\
        
        8 & Edit Fitness Goals &
        A screen where users set or adjust personal targets for
        calories, steps, or sleep. \\
        
        9 & Notifications &
        A screen showing alerts about upcoming workouts, reminders,
        or gym promotions. \\
        
        10 & Workout Goals &
        A screen where users view and track progress toward their
        set workout objectives. \\
        
        \bottomrule
    \end{tabular}

    \Description{This table lists screen keywords and evaluation prompts for Set 1, Session 2, which focuses on a Fitness Tracking App. It details ten specific screen types, ranging from an "Achievements Dashboard" and "Exercise Instruction" to "Notifications" and "Workout Goals." Each entry pairs a keyword with a descriptive prompt outlining the expected content for that screen in the evaluation arena.}
\end{table}


\subsection{Set 2}
\label{sec:set2}


\subsubsection{Session 1: Food Delivery App}
\label{sec:set2-session1}

\paragraph{Screen Keywords Given to Participants.}
\begin{itemize}
    \item Alerts
    \item Restaurant Bookmark
    \item Cart
    \item Menu Detail
\end{itemize}

\paragraph{Prompts Used in the Evaluation Arena.}
\begin{table}[H]
    \centering
    \caption{
        Screen keywords and prompts used in the evaluation arena
        for Set 2, Session 1 (Food Delivery App).
    }
    \label{tab:set2-session1-arena}
    \begin{tabular}{c p{0.25\textwidth} p{0.6\textwidth}}
        \toprule
        \textbf{Index} &
        \textbf{Screen Keyword} &
        \textbf{Prompt} \\
        \midrule
        1 & Alerts &
        A screen where users can view notifications about order
        updates, promotions, and app announcements. \\
        
        2 & Restaurant Bookmark &
        A screen where users can see and manage the list of
        restaurants and menu items they've saved. \\
        
        3 & Order Detail &
        A screen where users can review the full breakdown of a
        past or current order. \\
        
        4 & Checkout &
        A screen where users can confirm delivery address, payment,
        and order summary before placing an order. \\
        
        5 & Restaurant Detail &
        A screen where users can browse a restaurant's info,
        ratings, and menu. \\
        
        6 & Address Edit &
        A screen where users can see the current preferred delivery
        address and edit saved addresses. \\
        
        7 & Chat screen with delivery driver &
        A screen where users can message their assigned delivery
        driver in real time. \\
        
        8 & Restaurant review &
        A screen where users can write and submit a review for a
        restaurant after an order. \\
        
        9 & Order custom &
        A screen where users can customize a menu item's options,
        add-ons, and quantity before adding it to cart. \\
        
        10 & Profile &
        A screen where users can view and manage their account info
        and settings. \\
        
        \bottomrule
    \end{tabular}
    \Description{This table presents screen keywords and prompts for Set 2, Session 1, dedicated to a Food Delivery App. It lists ten specific screen types, such as "Restaurant Bookmark," "Checkout," and "Chat screen with delivery driver". Each entry associates a keyword with a detailed prompt defining the expected layout and functionality for that screen in the evaluation arena.}
\end{table}


\subsubsection{Session 2: Digital Payments App}
\label{sec:set2-session2}

\paragraph{Screen Keywords Given to Participants.}
\begin{itemize}
    \item Linked Card List
    \item Internet Bill Dashboard
    \item Transaction History
    \item Contact
\end{itemize}

\paragraph{Prompts Used in the Evaluation Arena.}
\begin{table}[H]
    \centering
    \caption{
        Screen keywords and prompts used in the evaluation arena
        for Set 2, Session 2 (Digital Payments App).
    }
    \label{tab:set2-session2-arena}
    \begin{tabular}{c p{0.25\textwidth} p{0.6\textwidth}}
        \toprule
        \textbf{Index} &
        \textbf{Screen Keyword} &
        \textbf{Prompt} \\
        \midrule
        1 & Top Up &
        A screen where users add money to their wallet from a linked
        bank account or card. \\
        
        2 & Linked Card List &
        A screen where users view and manage all bank cards and
        accounts connected to their wallet. \\
        
        3 & Electricity Bill Dashboard &
        A screen where users view their electricity account, usage,
        due amount, and payment history. \\
        
        4 & Transaction Detail &
        A screen where users see the full breakdown of a single
        past transaction. \\
        
        5 & Add New Contact &
        A screen where users manually enter and save a new
        recipient's details for future transfers. \\
        
        6 & Split Money Request with Friends &
        A screen where users divide a bill among multiple contacts
        and send each a payment request. \\
        
        7 & QR Scan Payment &
        A screen where users scan a merchant's QR code to make
        a payment. \\
        
        8 & Login &
        A screen where users enter their credentials or use
        biometrics to access their account. \\
        
        9 & Savings with Goal &
        A screen where users set up and track a savings goal with
        a target amount and progress. \\
        
        10 & Payment Receipt &
        A screen confirming a completed payment with amount,
        recipient, and transaction details. \\
        
        \bottomrule
    \end{tabular}
    \Description{This table presents screen keywords and prompts for Set 2, Session 2, dedicated to a Digital Payments App. It lists ten specific screen types, such as "Top Up," "Electricity Bill Dashboard," and "Split Money Request with Friends". Each entry associates a keyword with a detailed prompt defining the expected layout and functionality for that screen in the evaluation arena.}
\end{table}


\subsection{Interview Questionnaire}
\label{sec:interview-questionnaire}

\begin{enumerate}
    \item How was your overall experience using the plugin?
    \item Did you feel any inconvenience or difficulty using the plugin?
    Or was there any situation where you felt something was unnatural?
    \item What criteria did you use to pick one screen from the 8
    candidates?
    \item Did you feel that the model's generations became more
    conformant with each round?
    \begin{enumerate}
        \item Rate from 1 (strongly disagree) to 10 (strongly agree),
        where ``strongly agree'' means you didn't need to provide
        additional feedback to the model.
    \end{enumerate}
        \item If you could give the model other types of feedback besides
    direct modification, what type would you want, and why?
      \item Imagine that you are a designer at a company and will use this system for a genUI product. What is your main concern, and what other feature/mechanisms do you think are needed?
\end{enumerate}

\end{document}